\documentclass[12pt,oneside,reqno]{amsart}
\usepackage[latin9]{inputenc}
\usepackage{geometry}
\usepackage{mathrsfs}
\usepackage{mathtools}
\usepackage{amstext}
\usepackage{amsthm}
\usepackage{amssymb}
\usepackage{mathdots}
\usepackage{graphicx}
\usepackage{esint}
\usepackage[unicode=true,pdfusetitle,
 bookmarks=true,bookmarksnumbered=true,bookmarksopen=true,bookmarksopenlevel=2,
 breaklinks=false,pdfborder={0 0 1},backref=section,colorlinks=false]
 {hyperref}
\hypersetup{
 hypertex,bookmarksnumbered=true}

\makeatletter

\DeclareTextSymbolDefault{\textquotedbl}{T1}
\providecommand{\tabularnewline}{\\}

\numberwithin{equation}{section}
\numberwithin{figure}{section}
\theoremstyle{plain}
\newtheorem{thm}{\protect\theoremname}
\theoremstyle{definition}
\newtheorem{defn}[thm]{\protect\definitionname}
\theoremstyle{remark}
\newtheorem{rem}[thm]{\protect\remarkname}

\usepackage{chngcntr}
\counterwithout{figure}{section}
\makeatother

\providecommand{\definitionname}{Definition}
\providecommand{\remarkname}{Remark}
\providecommand{\theoremname}{Theorem}

\begin{document}
\title{Simple Reciprocal Electric Circuit Exhibiting Exceptional Point of
Degeneracy}
\author{Kasra Rouhi}
\author{Filippo Capolino}
\author{Alexander Figotin}
\address{University of California at Irvine, CA 92967}
\email{afigotin@uci.edu}
\begin{abstract}
An exceptional point of degeneracy (EPD) occurs when both the eigenvalues
and the corresponding eigenvectors of a square matrix coincide and
the matrix has a nontrivial Jordan block structure. It is not easy
to achieve an EPD exactly. In our prior studies, we synthesized simple
conservative (lossless) circuits with evolution matrices featuring
EPDs by using two LC loops coupled by a gyrator. In this paper, we
advance even a simpler circuit with an EPD consisting of only two
LC loops with one capacitor shared. Consequently, this circuit involves
only four elements and it is perfectly reciprocal. The shared capacitance
and parallel inductance are negative with values determined by explicit
formulas which lead to EPD. This circuit can have the same Jordan
canonical form as the nonreciprocal circuit we introduced before.
This implies that the Jordan canonical form does not necessarily manifest
systems' nonreciprocity. It is natural to ask how nonreciprocity is
manifested in the system's spectral data. Our analysis of this issue
shows that nonreciprocity is manifested explicitly in: (i) the circuit
Lagrangian and (ii) the breakdown of certain symmetries in the set
of eigenmodes. All our significant theoretical findings were thoroughly
tested and confirmed by extensive numerical simulations using commercial
circuit simulator software.
\end{abstract}

\maketitle

\section{Introduction\label{sec:intro}}

A key motivation for this work is an interest in systems that exhibit
exceptional points of degeneracy (EPDs). EPD refers to the degeneracy
of the system matrix that occurs when both the eigenvalues and the
corresponding eigenvectors of the system coincide \cite{kato1966perturbation,heiss1990avoided,heiss2004exceptional,stehmann2004observation,heiss2012physics}.
The corresponding system matrix is not diagonalizable at EPD. Also,
EPD occurs when the system matrix is similar to a matrix containing
a nontrivial Jordan block that is a Jordan block with a size of at
least two \cite{heiss2004exceptionalJPhysA}. One application of EPD
systems is high-sensitivity, which has attracted a great deal of interest
\cite{wiersig2014enhancing,wiersig2016sensors,chen2017exceptional}.

Considerable efforts are required to design an EPD system, and several
methods have been proposed for achieving EPD. Those approaches are
based on: (i) non-Hermitian parity-time (PT) symmetric coupled systems
with balanced loss and gain \cite{bender1998real,ramezani2010unidirectional,barashenkov2015dimer};
(ii) lossless and gainless structures associated with a stationary
inflection point (SIP) and degenerate band edge (DBE) \cite{othman2015demonstration,nada2017theory,nada2020frozen,mealy2020general,herrero2022frozen};
(iii) coupled resonators \cite{schindler2011experimental,schindler2012PTsymmetric,hodaei2017enhanced};
and (iv) time-varying systems \cite{kazemi2019exceptional,rouhi2020exceptional,kazemi2022experimental,nikzamir2023time}.
Additionally, the EPD is investigated in a nonreciprocal circuit consisting
of two LC resonators without gain and loss coupled via a nonreciprocal
element, i.e., a gyrator \cite{figotin2020synthesis,figotin2021perturbations,nikzamir2021demonstration,rouhi2022exceptional,rouhi2022high}.
Although one of the approaches suggests that loss and gain are essential
for EPD, our recent studies indicate that an EPD can be obtained in
time-varying \cite{kazemi2019exceptional,rouhi2020exceptional,kazemi2022experimental,nikzamir2023time}
and gyrator-based \cite{figotin2020synthesis,figotin2021perturbations,nikzamir2021demonstration,rouhi2022exceptional,rouhi2022high}
systems without gain and losses.

\begin{figure}
\begin{centering}
\includegraphics[width=0.85\textwidth]{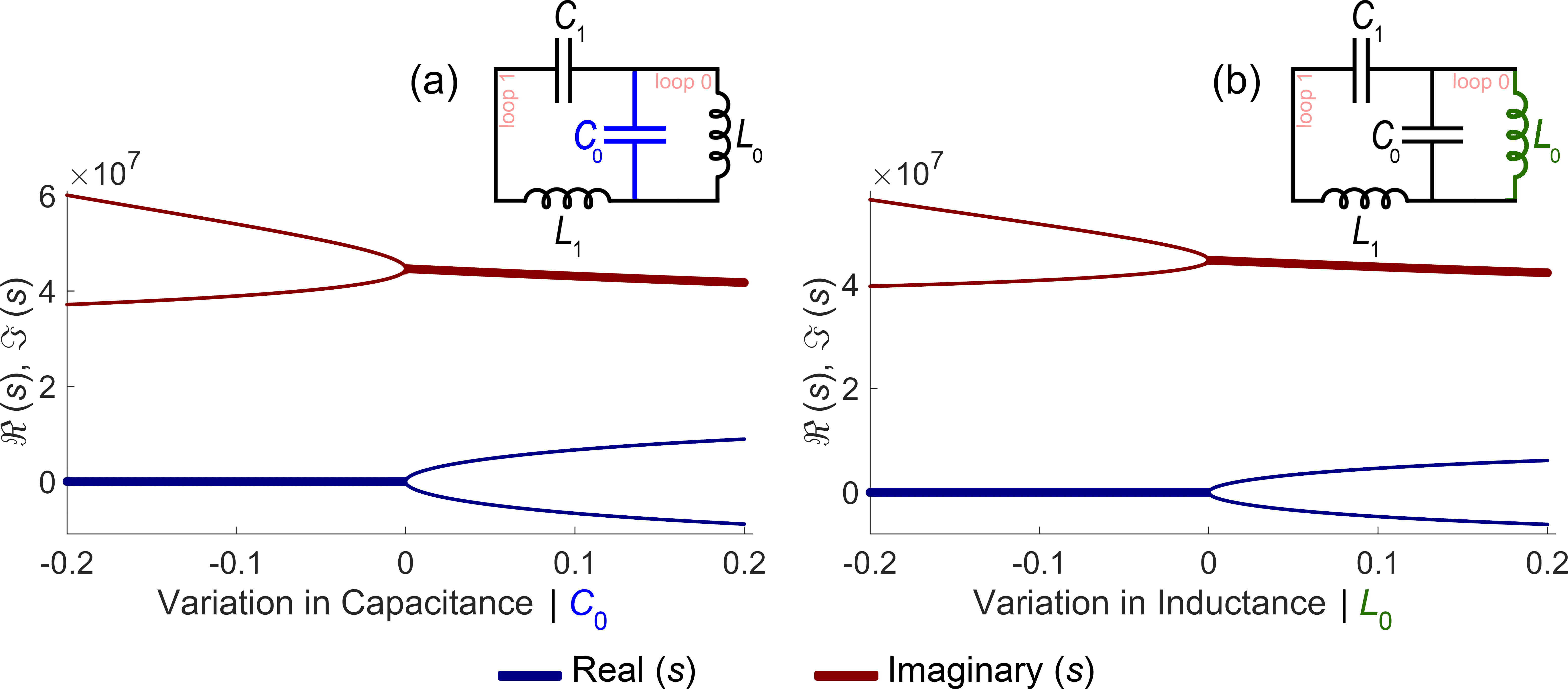}
\par\end{centering}
\caption{The real (dark blue solid curve) and imaginary (dark red solid curve)
parts of the eigenvalues ($s=\mathrm{i}\omega$, where $\omega$ is
eigenfrequency) of the circuit by varying (a) capacitance ($C_{0}$)
and (b) inductance ($L_{0})$. In these plots, thicker lines indicate
a multiplicity of two, and we limited the plots to eigenvalues with
positive imaginary parts. Due to the bifurcation near the EPD, a second-order
EPD with high sensitivity can be applied to sensing applications.
Also, the proposed reciprocal circuit is shown here where the perturbed
elements are colored in blue (capacitor) and green (inductor).\label{fig:DispersionIntro}}
\end{figure}

\textit{Following studies in \cite{figotin2020synthesis,figotin2021perturbations,nikzamir2021demonstration,rouhi2022exceptional,rouhi2022high}
we ask whether conservative and reciprocal circuits exist such that
their evolution matrices exhibit EPDs. The answer to this question
is positive and we construct here a circuit with EPD that does not
involve nonreciprocal or lossy elements (see circuits in Figure \ref{fig:DispersionIntro}).}
One of our primary goals here is to synthesize a conservative circuit
by using only reciprocal components so that its evolution matrix has
the nontrivial Jordan canonical form subject to natural constraints
considered later on as will be discussed in Section \ref{subsec:Characteristic-polynomial-and}.

In this paper, we advance a perfectly conservative and reciprocal
circuit that can be used to obtain the desired Jordan canonical form
and achieve second-order EPD. Our comparative studies of reciprocal
and nonreciprocal circuits suggest that circuit reciprocity information
is not necessarily encoded in the relevant Jordan canonical form but
rather in certain symmetries of the system eigenvectors. In addition,
both reciprocal and nonreciprocal schemes can lead to the same eigenvalues
whereas their Lagrangian equations are different. We demonstrate the
conditions for equivalence between reciprocal and nonreciprocal circuits
in general (i.e., when circuits are degenerate or non-degenerate)
to have the same Jordan canonical form despite different Lagrangian
equations. The present study includes mathematical analysis as well
as extensive time-domain numerical simulation results for verification
that were computed by a well-known commercial circuit simulator. The
eigenvalues of the proposed circuit, which will be discussed later,
are exceedingly sensitive to perturbations in circuit parameters (such
as capacitance or inductance) as shown in Figure \ref{fig:DispersionIntro}.
Hence, the proposed circuit provides exceptional capabilities for
applications that require high-sensitivity sensing when a component
of the circuit, which is essentially a sensing component, is perturbed
externally.

The structure of the paper is as follows. The mathematical setup of
the problem is presented in Section \ref{sec:Mathematical-setup-of}.
The main achievements and results of this paper are summarized in
Section \ref{sec:main}. Then, we demonstrate our primary circuit
with lossless and reciprocal elements in Section \ref{sec:Principal-Reciprocal-Circuit}.
We study the Jordan canonical form of the circuit and the condition
to obtain second-order EPD in our proposed circuit in Section \ref{sec:Jordan-Form}.
Section \ref{sec:Lagrangian-and-Hamiltonian} demonstrates the Lagrangian
and Hamiltonian structures and their relation in the general case.
Next, we review briefly the gyrator-based circuit introduced in our
previous studies and explain the mathematical analysis behind the
nonreciprocal design in Section \ref{sec:Gyrator-based-Circuit}.
In Section \ref{sec:Analysis-of-Reciprocity}, we analyze both reciprocal
and nonreciprocal circuits to identify the signs of nonreciprocity
in the Lagrangian equations and eigenvectors of the circuits. Also,
we provide the equivalency condition for both reciprocal and nonreciprocal
circuits in Section \ref{sec:Equivalency} to have the same eigenvalues
while the circuits' Lagrangian equations are different. Finally, we
support our mathematical analysis and findings by giving examples
that are verified by using a time-domain circuit simulator in Section
\ref{sec:Circuit-Simulator-Simulation} and wrapping up the paper
in Section \ref{sec:Conclusions}. Also, we include many appendices
for readers to provide more information and details.

\section{Mathematical Setup of The Problem\label{sec:Mathematical-setup-of}}

Our primary goal here is to develop a lossless electric circuit with
$2$ \emph{fundamental loops} (see circuit in Figure \ref{fig:DispersionIntro})
so that its evolution matrix $\mathscr{H}$ has a prescribed Jordan
canonical form $\mathscr{J}$ subject to natural constraints considered
later on. According to the definition of the Jordan canonical form,
we have $\mathscr{H}=\mathscr{S}\mathscr{J}\mathscr{S}^{-1}$ where
$\mathscr{S}$ is an invertible $4\times4$ matrix for the two-loop
circuit that is a block diagonal matrix of the form

\begin{gather}
\mathscr{J}=\left[\begin{array}{cc}
J\left(\zeta_{1}\right) & 0_{2}\\
0_{2} & J\left(\zeta_{2}\right)
\end{array}\right],\quad J\left(\zeta_{j}\right)=\left[\begin{array}{cc}
\zeta_{j} & 1\\
0 & \zeta_{j}
\end{array}\right],\quad j=1,2,\label{eq:Joblo1b}
\end{gather}
where $0_{2}$ is $2\times2$ zero matrix, $\zeta_{j}$ are real or
complex numbers, and $J\left(\zeta_{j}\right)$ are the corresponding
so-called Jordan block. As to the evolution circuit matrix $\mathscr{H}$
we assume that the circuit evolution is governed by the following
Hamilton evolution equation
\begin{equation}
\partial_{t}\mathscr{X}=\mathscr{H}\mathscr{X},\label{eq:XHX1a}
\end{equation}
where $\mathscr{X}$ is $4$ dimensional vector-column describing
the circuit state and $\mathscr{H}$ is $4\times4$ matrix. The matrix
$\mathscr{H}$ is going to be a Hamiltonian matrix and we will refer
to it as the \emph{circuit evolution matrix} or just \emph{circuit
matrix} (see Section \ref{sec:Lagrangian-and-Hamiltonian}). Also,
the circuit state vector is described by the corresponding two time
dependent charges $q_{j}\left(t\right)$ ($j=1,2$) which are the
time integrals of the relevant loop currents $\partial_{t}q_{k}\left(t\right)$
(see the circuits in Figure \ref{fig:DispersionIntro}).

The eigenvalue problem of Equation (\ref{eq:XHX1a}) can be written
as
\begin{equation}
s\mathscr{X}=\mathscr{H}\mathscr{X},\quad s=\mathrm{i}\omega,\label{eq:XHX1aa}
\end{equation}
where $s$ is the eigenvalue and $\omega$ is the eigenfrequency.
It turns out that if the Jordan canonical form $\mathscr{J}$ of the
circuit matrix $\mathscr{H}$ has a nontrivial Jordan block then the
circuit must have at least one negative capacitance and inductance
(see Section \ref{subsec:Characteristic-polynomial-and}). The origin
of the constraints is the fundamental property of a Hamiltonian matrix
$\mathscr{H}$ to be similar to $-\mathscr{H}^{\mathrm{T}}$ where
$\mathscr{H}^{\mathrm{T}}$ is the transposed to $\mathscr{H}$ matrix.
Apart from the Hamiltonian spectral symmetry the Jordan structure
of Hamiltonian matrices can be arbitrary. Our approach to the generation
of Hamiltonian and the corresponding Hamiltonian matrices is intimately
related to the Hamiltonian canonical forms (see Appendix F of \cite{figotin2020synthesis}).

Another significant mathematical input to the synthesis of the simplest
possible systems exhibiting nontrivial Jordan blocks comes from the
property of a square matrix $\mathscr{M}$ to be \emph{cyclic} (also
called \emph{non-derogatory}) (see Appendix B of \cite{figotin2020synthesis}).
We remind that a square matrix $\mathscr{M}$ is called cyclic if
the geometric multiplicity of each of its eigenvalues is exactly $1$.
It means that every eigenvalue of matrix $\mathscr{M}$ has exactly
one eigenvector. Consequently, if a square matrix $\mathscr{M}$ is
cyclic its Jordan form $\mathscr{J}_{\mathscr{M}}$ is completely
determined by its \emph{characteristic polynomial} $\chi\left(s\right)=\det\left\{ s\mathbb{I}_{4}-\mathscr{M}\right\} $
where $\mathbb{I}_{4}$ is the $4\times4$ identity matrix. Namely,
every eigenvalue $s_{0}$ of $\mathscr{M}$ of multiplicity $2$ is
associated with the single Jordan block $J\left(s_{0}\right)$ in
the Jordan form $\mathscr{J}_{\mathscr{M}}$ of $\mathscr{M}$. \textit{Consequently,}
\emph{for a cyclic matrix $\mathscr{M}$ its characteristic polynomial
encodes all the information about its Jordan form} $\mathscr{J}_{\mathscr{M}}$.
Another property of any cyclic matrix $\mathscr{M}$ associated with
the a monic polynomial $\chi$ is that it is similar to the so-called
companion matrix $C_{\chi}$ defined by simple explicit expression
involving the coefficients of the polynomial $\chi$ (see Appendix
B of \cite{figotin2020synthesis}).

Companion matrix $C_{\chi}$ is naturally related to the high-order
differential equation $\chi\left(\partial_{t}\right)x\left(t\right)=0$,
where $x\left(t\right)$ is a complex-valued function of $t$ (see
Appendices B and D of \cite{figotin2020synthesis}). This fact underlines
the relevance of the cyclicity property to the evolution of simpler
systems described by higher order differential equations for a scalar
function. Accordingly, we focus on cyclic Hamiltonian matrices $\mathscr{H}$
for they lead to the simplest circuits with the evolution matrices
$\mathscr{H}$ having nontrivial Jordan forms $\mathscr{J}$.

Suppose the prescribed Jordan form $\mathscr{J}$ is an $4$$\times4$
matrix subject to the Hamiltonian spectral symmetry and the cyclicity
conditions. For this circuit we have the characteristic polynomial
$\chi\left(s\right)=\det\left\{ s\mathbb{I}_{4}-\mathscr{J}\right\} $
which is an even monic polynomial $\chi\left(s\right)$ of the degree
4. We consider then the companion to $\chi\left(s\right)$ matrix
$\mathscr{C}$ (see Appendix B of \cite{figotin2020synthesis}), that
is

\begin{equation}
\mathscr{C}=\mathscr{Y}\mathscr{J}\mathscr{Y}^{-1},\label{eq:XHX1b}
\end{equation}
where the columns of matrix $\mathscr{Y}$ form the so-called Jordan
basis of the companion matrix $\mathscr{C}$ associated with the characteristic
polynomial $\chi\left(s\right)=\det\left\{ s\mathbb{I}_{4}-\mathscr{J}\right\} $.
We proceed with an introduction of our principal Hamiltonian $\mathcal{H}$
and recover from it $4\times4$ Hamiltonian matrix $\mathscr{H}$
that governs the system evolution according to Equation (\ref{eq:XHX1a}).
As the result of our particular choice of the Hamiltonian $\mathcal{H}$
the corresponding to it Hamiltonian matrix $\mathscr{H}$ is similar
to the companion matrix $\mathscr{C}$ and consequently it has exactly
the same Jordan form $\mathscr{J}$ as $\mathscr{C}$. In particular,
we construct an $4\times4$ matrix $\mathscr{T}$ such that
\begin{equation}
\mathscr{C}=\mathscr{T}^{-1}\mathscr{H}\mathscr{T},\quad\mathscr{H}=\mathscr{Z}\mathscr{J}\mathscr{Z}^{-1},\quad\mathscr{Z}=\mathscr{T}\mathscr{Y},\label{eq:XHX1c}
\end{equation}
where the columns of matrix $\mathscr{Z}$ form a Jordan basis of
the evolution matrix $\mathscr{H}$. The Equations (\ref{eq:XHX1b})
and (\ref{eq:XHX1c}) between involved matrices are considered in
Section \ref{sec:Lagrangian-and-Hamiltonian}.

\section{Review of The Main Results\label{sec:main}}

We advance here a perfectly conservative circuit without a gyrator
which has an EPD and consequently a nontrivial Jordan canonical form.
This circuit is shown in Figure \ref{fig:TwoCircuitCompare}(a) and
we will refer to it as a \textit{principle reciprocal circuit} (PRC)
or simply a \textit{reciprocal circuit} in the rest of the paper.
The circuit has only four elements which are all lossless and reciprocal.
The nontrivial Jordan canonical form is already obtained in the \textit{gyrator-based
nonreciprocal circuit} (GNC) which has a nonreciprocal element \cite{figotin2020synthesis,figotin2021perturbations,nikzamir2021demonstration,nikzamir2022achieve,rouhi2022high},
as shown in Figure \ref{fig:TwoCircuitCompare}(b). The list of components
for reciprocal and nonreciprocal circuits and the necessary conditions
to pick the equivalency value of gyration resistance is summarized
in Table \ref{tab:Summary}. The list of our main achievements that
are elaborated in the rest of the paper is as follows:

\begin{figure}[h]
\begin{centering}
\includegraphics[width=0.7\textwidth]{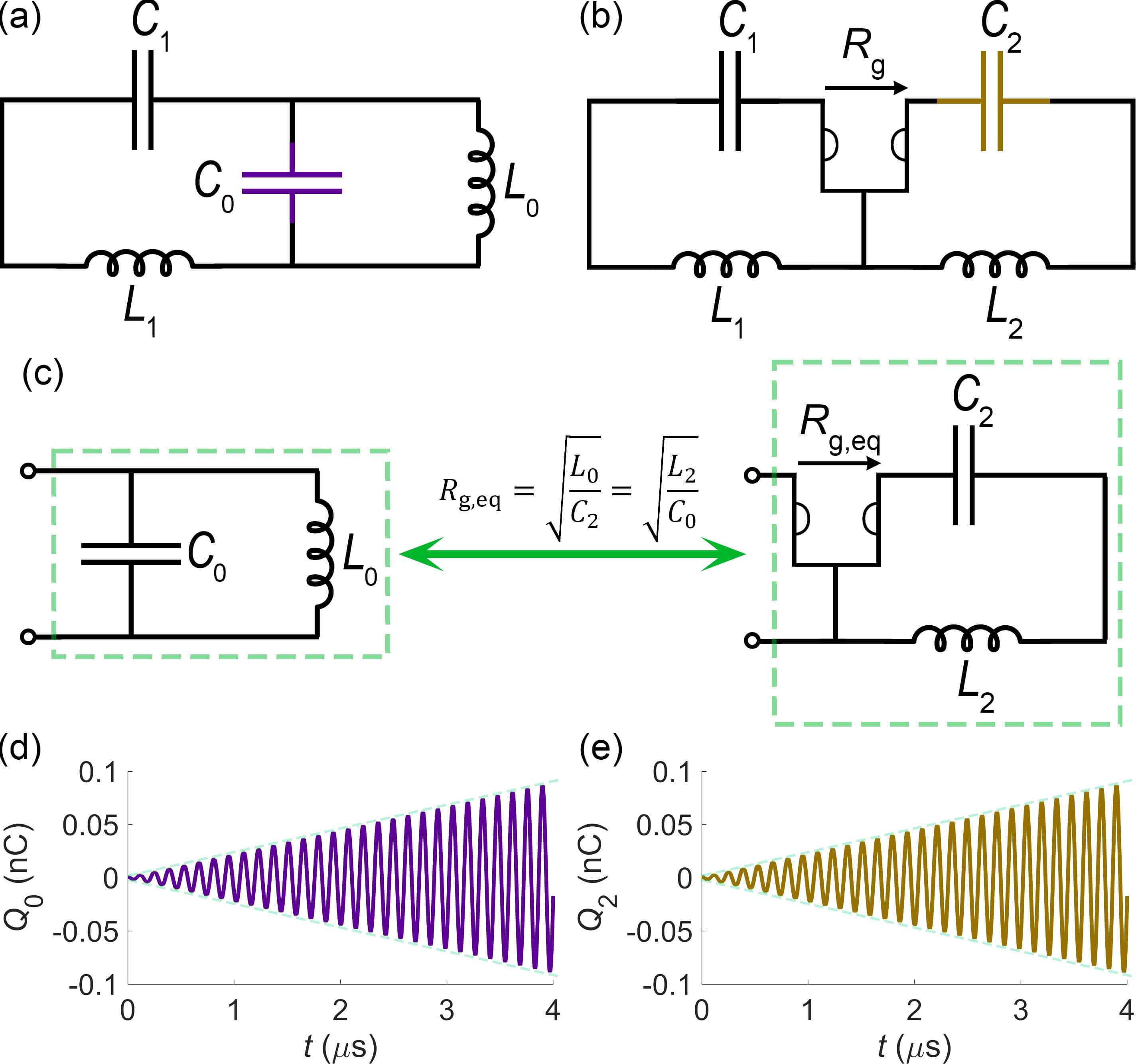}
\par\end{centering}
\caption{Comparison between (a) the 4-element reciprocal circuit (the focus
of this paper), also denoted as PRC, and (b) the 5-element nonreciprocal
circuit that uses a gyrator as a nonreciprocal element, also denoted
as GNC. (c) Equivalent resonators with the equivalency value of gyration
resistance $R_{\mathrm{g,eq}}=\sqrt{L_{0}/C_{2}}=\sqrt{L_{2}/C_{0}}$.
Stored charges calculated by time-domain circuit simulator in (d)
the capacitor of the PRC $C_{0}$ and (e) the capacitor of the GNC
$C_{2}\left(t\right)$ under the both EPD and equivalency conditions.
Furthermore, the envelope of the charge signals shows linear growth
over time, which is a distinguishing characteristic of second-order
EPDs.\label{fig:TwoCircuitCompare}}
\end{figure}

\begin{itemize}
\item \textit{Construction of a reciprocal and lossless circuit with an
EPD.} We prove that the nontrivial Jordan canonical form can be obtained
in a circuit without gyrator (see Section \ref{sec:Principal-Reciprocal-Circuit}).
\item \textit{PRC can have exactly the same Jordan canonical form as GNC
and we find conditions for this to occur.} If we consider the gyration
resistance value of $R_{\mathrm{g,eq}}$ (see Figure \ref{fig:TwoCircuitCompare}(c)),
the reciprocal and nonreciprocal circuits can have the same Jordan
canonical forms implying that the eigenvalues of both circuits are
the same. We conducted extensive numerical simulations of PRC and
GNC using a commercial time-domain circuit simulator for comparison.
In particular, we observe the charge stored in the capacitor associated
with right loops as shown in Figures \ref{fig:TwoCircuitCompare}(d)
and (e) for reciprocal ($C_{0}$) and nonreciprocal ($C_{2}$) circuits,
respectively. There is excellent agreement between numerical simulations
and the above condition of equivalency is satisfied as demonstrated
by numerical simulations (see Sections \ref{sec:Circuit-Simulator-Simulation}
and \ref{sec:Equivalency}).
\item \textit{Our analysis of the spectral data shows that nonreciprocity
in the GNC is manifested in the breakdown of certain symmetries for
the set of eigenvectors while this symmetry exists for PRC.} The nonreciprocity
is also manifested in the circuit Lagrangian. Despite this, nonreciprocity
is not captured by analyzing the eigenvalues or the Jordan canonical
form of the circuit matrix (see Section \ref{sec:Analysis-of-Reciprocity}).
\end{itemize}
\begin{table}
\caption{List of component and required conditions for equivalent reciprocal
and nonreciprocal circuits.\label{tab:Summary}}

\begin{centering}
\begin{tabular}{|r|r|r|r|}
\hline 
 & 4-Element$^{\star}$ & 5-Element$^{\dagger}$ & Equivalency\tabularnewline
\hline 
\hline 
Capacitance - first loop & $C_{1}$ & $C_{1}$ & \tabularnewline
\cline{1-3} \cline{2-3} \cline{3-3} 
Inductance - first loop & $L_{1}$ & $L_{1}$ & $L_{1}C_{1}=L_{1}C_{1}$\tabularnewline
\hline 
Capacitance - second loop & $C_{0}$ & $C_{2}$ & \tabularnewline
\cline{1-3} \cline{2-3} \cline{3-3} 
Inductance - second loop & $L_{0}$ & $L_{2}$ & $L_{2}C_{2}=L_{0}C_{0}$\tabularnewline
\hline 
Gyrator & None & $R_{g}$ & $R_{\mathrm{g,eq}}=\sqrt{\frac{L_{0}}{C_{2}}}=\sqrt{\frac{L_{2}}{C_{0}}}$\tabularnewline
\hline 
\end{tabular}
\par\end{centering}
{\footnotesize{}$\,$}{\footnotesize\par}
\begin{raggedright}
{\footnotesize{}$\qquad{}^{\star}$ 4-Element: Principal reciprocal
circuit (PRC)}{\footnotesize\par}
\par\end{raggedright}
\raggedright{}{\footnotesize{}$\qquad{}^{\dagger}$ 5-Element: Gyrator-based
nonreciprocal circuit (GNC)}{\footnotesize\par}
\end{table}

\section{Principal Reciprocal Circuit (PRC)\label{sec:Principal-Reciprocal-Circuit}}

\begin{figure}
\begin{centering}
\includegraphics[width=0.35\textwidth]{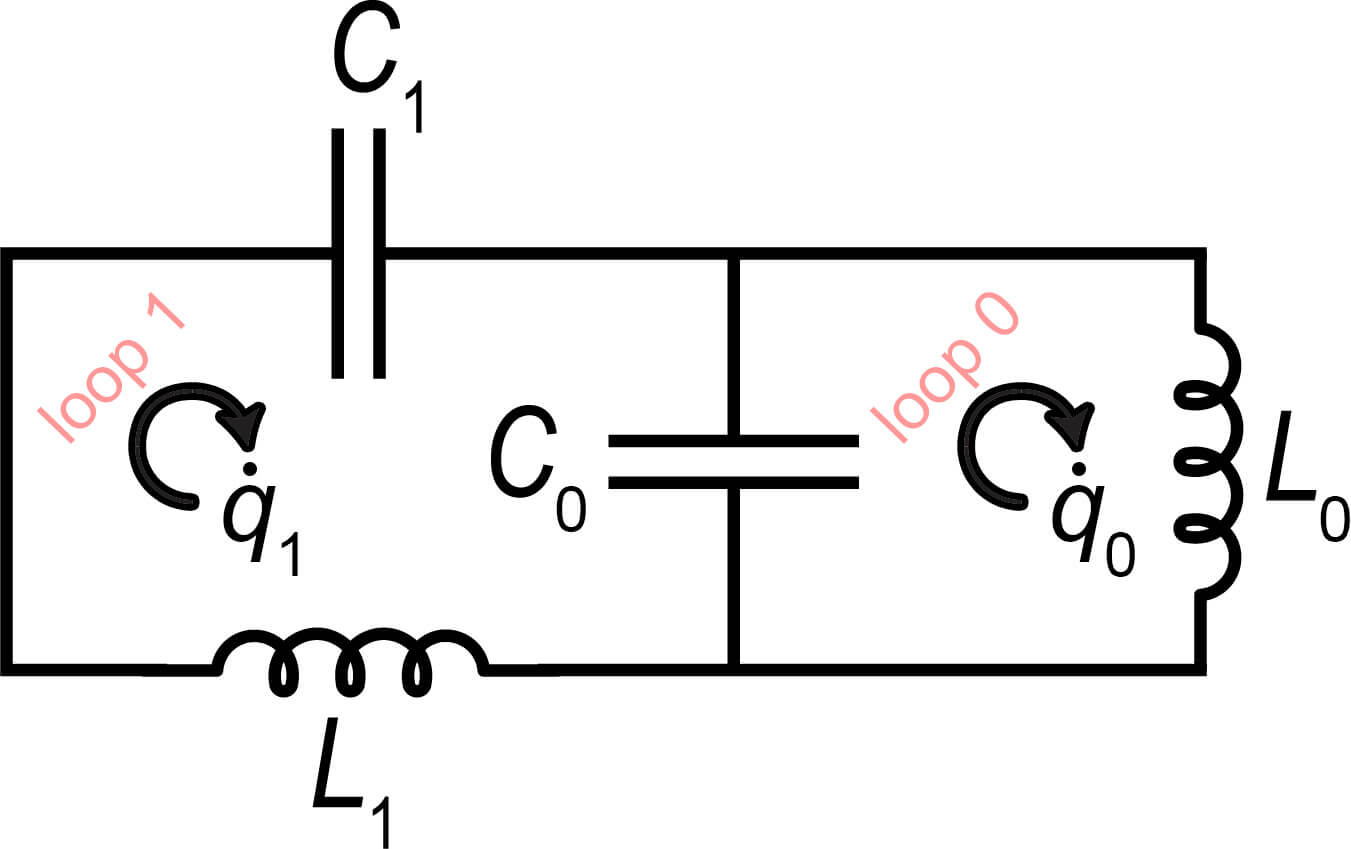}
\par\end{centering}
\caption{The principal two-loop circuit with reciprocal and lossless elements.
For particular choices of values for quantities $L_{1}$, $C_{1}$,
$L_{0}$, and $C_{0}$, the evolution matrix of the circuit develops
second-order degeneracy, and its nontrivial Jordan canonical form
consists of exactly two Jordan blocks of size 2. In this circuit,
it is possible to replace $C_{0}$ and $L_{0}$ with each other since
they are parallel. It is noteworthy that the zeroth loop and the first
loop serve different purposes, with the capacitance $C_{0}$ being
shared within the loops to couple them together.\label{fig:RecipCir}}
\end{figure}
We advance here our proposed simple reciprocal circuit with a circuit
matrix having the nontrivial Jordan canonical form without the need
for gyrator. Figure \ref{fig:RecipCir} shows the PRC made of two
fundamental loops connected directly. In this circuit, quantities
$L_{j}$ and $C_{j}$ ($j=0,1$) are respectively inductance and capacitance
of the corresponding loops. The\textit{ Lagrangian} associated with
PRC displayed in Figure \ref{fig:RecipCir} is expressed by

\begin{subequations}
\begin{equation}
\left.\mathcal{L}_{\mathrm{r}}\left(\mathrm{q}_{\mathrm{r}},\partial_{t}\mathrm{q}_{\mathrm{r}}\right)\right|_{\Theta_{\mathrm{r}}}=\frac{L_{1}\left(\partial_{t}q_{1}\right)^{2}}{2}+\frac{L_{0}\left(\partial_{t}q_{0}\right)^{2}}{2}-\frac{\left(q_{1}\right)^{2}}{2C_{1}}-\frac{\left(q_{1}-q_{0}\right)^{2}}{2C_{0}},\label{eq:RecipLag}
\end{equation}

\begin{equation}
q_{j}\left(t\right)=\int^{t}i_{j}\left(t'\right)\,\mathrm{d}t',\quad j=0,1,
\end{equation}
\end{subequations}
where $q_{j}$ and $i_{j}$ are the charges and the currents associated
with loops of the PRC depicted in Figure \ref{fig:RecipCir} and $\Theta_{\mathrm{r}}=\left\{ L_{1},L_{0},C_{1},C_{0}\right\} $
is the set of circuit's parameters. We introduce a vector of charges
as $\mathrm{q}_{\mathrm{r}}=\left[q_{1},q_{0}\right]^{\mathrm{T}}$
($\mathrm{T}$ denotes the transpose operator) that composed of the
charges associated with two fundamental loops. According to time reversal
symmetry in reciprocal systems, the Lagrangian associated with the
PRC depicted in Figure \ref{fig:RecipCir} has the property $\mathcal{L}_{\mathrm{r}}\left(\mathrm{q}_{\mathrm{r}},\partial_{t}\mathrm{q}_{\mathrm{r}}\right)=\mathcal{L}_{\mathrm{r}}\left(\mathrm{q}_{\mathrm{r}},-\partial_{t}\mathrm{q}_{\mathrm{r}}\right)$.
The reciprocity principle and features of the Lagrangian formulations
will be further examined in Subsection \ref{subsec:ReciprocityLageq}.
The corresponding Euler-Lagrange equations of the PRC are given by

\begin{subequations}
\begin{equation}
\frac{d}{dt}\left(\frac{\partial}{\partial\dot{q}_{1}}\mathcal{L}\right)-\frac{\partial}{\partial q_{1}}\mathcal{L}=0,
\end{equation}

\begin{equation}
\frac{d}{dt}\left(\frac{\partial}{\partial\dot{q}_{0}}\mathcal{L}\right)-\frac{\partial}{\partial q_{0}}\mathcal{L}=0,
\end{equation}
\end{subequations}
which are simplified in the form of

\begin{subequations}
\begin{equation}
L_{1}\ddot{q}_{1}+\left(\frac{1}{C_{1}}+\frac{1}{C_{0}}\right)q_{1}-\frac{1}{C_{0}}q_{0}=0,\label{eq:ELRecip1}
\end{equation}

\begin{equation}
L_{0}\ddot{q}_{0}-\frac{1}{C_{0}}q_{1}+\frac{1}{C_{0}}q_{0}=0.\label{eq:ELRecip2}
\end{equation}
\end{subequations}
It is well known that the Euler-Lagrange formulations of Equations
(\ref{eq:ELRecip1}) and (\ref{eq:ELRecip2}) represent the Kirchhoff
voltage law for each of the two fundamental loops. The Kirchhoff voltage
equations for the PRC are calculated in Appendix \ref{sec:Kirchoff's-Equations}.
Accordingly, each term in Equations (\ref{eq:ELRecip1}) and (\ref{eq:ELRecip2})
corresponds to the voltage drop of the relevant element, as can be
seen from the current-voltage relations reviewed in Appendix \ref{subsec:cir-elem}.

The circuit vector evolution equation by using the state vector of
$\mathsf{q}_{\mathrm{r}}=\left[\mathrm{q}_{\mathrm{r}},\partial_{t}\mathrm{q}_{\mathrm{r}}\right]^{\mathrm{T}}$
is given by

\begin{equation}
\partial_{t}\mathsf{q}_{\mathrm{r}}=\mathscr{C}_{\mathrm{r}}\mathsf{q}_{\mathrm{r}},\quad\mathscr{C}_{\mathrm{r}}=\left[\begin{array}{rrcr}
0 & 0 & 1 & 0\\
0 & 0 & 0 & 1\\
-\left(\frac{1}{L_{1}C_{1}}+\frac{1}{L_{1}C_{0}}\right) & \frac{1}{L_{1}C_{0}} & 0 & 0\\
\frac{1}{L_{0}C_{0}} & -\frac{1}{L_{0}C_{0}} & 0 & 0
\end{array}\right],\label{eq:RecipEigValProb}
\end{equation}
where $\mathscr{C}_{\mathrm{r}}$ is circuit matrix corresponding
to the PRC. The characteristic polynomial related to the matrix $\mathscr{C}_{\mathrm{r}}$
defined by Equation (\ref{eq:RecipEigValProb}) is expressed by

\begin{equation}
\chi_{\mathrm{r}}\left(s\right)=\det\left\{ s\mathbb{I}_{4}-\mathscr{C}_{\mathrm{r}}\right\} =s^{4}+s^{2}\left(\xi_{1}+\xi_{0}+\delta\right)+\xi_{1}\xi_{0}=0,\label{eq:RecipCharEq}
\end{equation}
where $\mathbb{I}_{4}$ is the $4\times4$ identity matrix and the
circuit resonance frequencies are defined as

\begin{equation}
\delta=\frac{1}{L_{1}C_{0}},\quad\xi_{0}=\frac{1}{L_{0}C_{0}},\quad\xi_{1}=\frac{1}{L_{1}C_{1}}.\label{eq:RecipPara}
\end{equation}
In the above equations, $\sqrt{\zeta_{0}}$ indicates the resonance
frequency of the zeroth loop, $\sqrt{\zeta_{1}}$ indicates the resonance
frequency of the first loop and $\sqrt{\delta}$ is defined as a cross
loop resonance frequency or coupling term. We demonstrate the following
properties in the structure of the characteristic polynomial: (i)
the characteristic polynomial is a quadratic equation in $s^{2}$,
so if $s$ is the solution of the characteristic polynomial then $-s$
is its solution as well; (ii) the coefficients of the characteristic
polynomial are real; hence $s$ and $s^{*}$ ($\mathrm{^{*}}$ denotes
the complex conjugate operation) are both solutions of the characteristic
polynomial. Note that using circuit resonance frequencies defined
in Equation (\ref{eq:RecipPara}), we recast the circuit matrix $\mathscr{C}_{\mathrm{r}}$
in Equation (\ref{eq:RecipEigValProb}) as below

\begin{equation}
\mathscr{C}_{\mathrm{r}}=\left[\begin{array}{rrcr}
0 & 0 & 1 & 0\\
0 & 0 & 0 & 1\\
-\xi_{1}-\delta & \delta & 0 & 0\\
\xi_{0} & -\xi_{0} & 0 & 0
\end{array}\right],\label{eq:RecipComp}
\end{equation}
which shows that matrix $\mathscr{C}_{\mathrm{r}}$ defined by Equation
(\ref{eq:RecipComp}) is a block off-diagonal matrix.

\section{The Jordan Canonical Form of PRC\label{sec:Jordan-Form}}

We studied the PRC composed of two loops, as shown in Figure \ref{fig:RecipCir},
without putting any constraints on the circuit parameters $L_{1}$,
$C_{1}$, $L_{0}$, and $C_{0}$ except that they are all real and
non-zero. In this section, we derive the most general conditions on
the circuit parameters under which the relevant evolution matrix shows
nontrivial Jordan blocks. The Lagrangian equation for the PRC is given
by Equation (\ref{eq:RecipLag}), and its evolution equations are
the corresponding Euler-Lagrange equations:

\begin{subequations}
\begin{equation}
s^{2}q_{1}+\left(\xi_{1}+\delta\right)q_{1}-\delta q_{0}=0,
\end{equation}

\begin{equation}
s^{2}q_{0}-\xi_{0}q_{1}+\xi_{0}q_{0}=0.
\end{equation}
\end{subequations}
The Euler-Lagrange equations are readily recast in the below matrix
form,

\begin{equation}
\mathrm{A}_{\mathrm{r}}\left(s\right)\mathrm{q}_{\mathrm{r}}=0,\quad\mathrm{A}_{\mathrm{r}}\left(s\right)=\left[\begin{array}{rr}
s^{2}+\xi_{1}+\delta & -\delta\\
-\xi_{0} & s^{2}+\xi_{0}
\end{array}\right],\label{eq:MatDiffEq}
\end{equation}
where $\mathrm{A}_{\mathrm{r}}\left(s\right)$ is a $2\times2$ monic
matrix polynomial of $s$ of degree 2, namely,

\begin{subequations}
\begin{equation}
\mathrm{A}_{\mathrm{r}}\left(s\right)=\sum_{j=0}^{2}s^{j}\mathrm{A}_{\mathrm{r},j}=s^{2}\mathrm{A}_{\mathrm{r},2}+s\mathrm{A}_{\mathrm{r},1}+\mathrm{A}_{\mathrm{r},0},\label{eq:MatDiffEq2}
\end{equation}

\begin{equation}
\mathrm{A}_{\mathrm{r},2}=\left[\begin{array}{rr}
1 & 0\\
0 & 1
\end{array}\right],\quad\mathrm{A}_{\mathrm{r},1}=\left[\begin{array}{rr}
0 & 0\\
0 & 0
\end{array}\right],\quad\mathrm{A}_{\mathrm{r},0}=\left[\begin{array}{rr}
\xi_{1}+\delta & -\delta\\
-\xi_{0} & \xi_{0}
\end{array}\right].
\end{equation}
\end{subequations}
Then, Equation (\ref{eq:MatDiffEq}) is reduced to the standard first-order
vector differential equation

\begin{equation}
s\mathsf{q}_{\mathrm{r,s}}=\mathscr{C}_{\mathrm{r}}\mathsf{q}_{\mathrm{r,s}},\quad\mathsf{q}_{\mathrm{r,s}}=\left[\begin{array}{r}
\mathrm{q}_{\mathrm{r}}\\
s\mathrm{q}_{\mathrm{r}}
\end{array}\right],\quad\mathscr{C}_{\mathrm{r}}=\left[\begin{array}{rrcr}
0 & 0 & 1 & 0\\
0 & 0 & 0 & 1\\
-\xi_{1}-\delta & \delta & 0 & 0\\
\xi_{0} & -\xi_{0} & 0 & 0
\end{array}\right],\label{eq:MatDiffEq4}
\end{equation}
where $\mathscr{C}_{\mathrm{r}}$ is the $4\times4$ companion matrix
for the matrix polynomial $\mathrm{A}_{\mathrm{r}}\left(s\right)$.
The standard eigenvalue problem is expressed as

\begin{equation}
\left(s\mathbb{I}_{4}-\mathscr{C}_{\mathrm{r}}\right)\mathsf{q}_{\mathrm{r,s}}=0.\label{eq:CharEqMatDiff}
\end{equation}
Then, the characteristic polynomial of the matrix polynomial $\mathrm{A}_{\mathrm{r}}\left(s\right)$
is

\begin{equation}
\chi_{\mathrm{r}}\left(s\right)=\det\left\{ \mathrm{A}_{\mathrm{r}}\left(s\right)\right\} =s^{4}+s^{2}\left(\xi_{1}+\xi_{0}+\delta\right)+\xi_{1}\xi_{0}.\label{eq:RecipCharEq2}
\end{equation}
Consequently, the characteristic polynomial $\chi_{\mathrm{r}}\left(s\right)=0$
can be used to calculate the eigenvalues associated with Equation
(\ref{eq:CharEqMatDiff}). We aim in this paper to find all non-zero
and real values of the circuit parameters $L_{1}$, $C_{1}$, $L_{0}$
and $C_{0}$ that lead to the matrix $\mathscr{C}_{\mathrm{r}}$ defined
by Equation (\ref{eq:MatDiffEq4}) having a nontrivial Jordan canonical
form.

\subsection{Characteristic polynomial and eigenvalue degeneracy\label{subsec:Characteristic-polynomial-and}}

We present here the condition for degenerate eigenvalues in the characteristic
polynomial. We rewrite the characteristic polynomial associated with
the PRC matrix as

\begin{equation}
\chi_{\mathrm{r}}\left(h\right)=h^{2}+h\left(\xi_{1}+\xi_{0}+\delta\right)+\xi_{1}\xi_{0}=0,\quad h=s^{2}.\label{eq:CharEqh}
\end{equation}
A quadratic equation in $h$, $\chi_{\mathrm{r}}\left(h\right)=0$,
has exactly two solutions, viz,

\begin{equation}
h_{\pm}=\frac{-\left(\xi_{1}+\xi_{0}+\delta\right)\pm\sqrt{\Delta_{\mathrm{r},h}}}{2},\label{eq:h+-}
\end{equation}
where $\Delta_{\mathrm{r},h}=\delta^{2}+2\delta\left(\xi_{1}+\xi_{0}\right)+\left(\xi_{1}-\xi_{0}\right)^{2}$
is the discriminant of the quadratic in the characteristic polynomial
of Equation (\ref{eq:CharEqh}). The four solutions of the characteristic
polynomial $\chi_{\mathrm{r}}\left(s\right)=0$ are as follows:

\begin{equation}
s=\pm\sqrt{h_{+}},\pm\sqrt{h_{-}}.
\end{equation}
Note that the eigenvalue degeneracy condition turns into $\Delta_{\mathrm{r},h}=0$,
which is equivalently expressed as

\begin{equation}
\delta^{2}+2\delta\left(\xi_{1}+\xi_{0}\right)+\left(\xi_{1}-\xi_{0}\right)^{2}=0.\label{eq:DegCon}
\end{equation}
Equation (\ref{eq:DegCon}) is a quadratic equation which has exactly
two solutions. We refer to solutions of $\delta$ in Equation (\ref{eq:DegCon})
as\emph{ special values of degeneracy $\delta_{\mathrm{d,\pm}}$,
}where these two solutions are

\begin{equation}
\delta_{\mathrm{d,\pm}}=-\xi_{1}-\xi_{0}\pm2\sqrt{\xi_{1}\xi_{0}}.\label{eq:SpecValDeg}
\end{equation}
For the two special values of degeneracy \emph{$\delta_{\mathrm{d,\pm}}$}
we get from Equation (\ref{eq:h+-}) the corresponding two degenerate
roots of $h$ are given by

\begin{equation}
h_{\mathrm{d,}\pm}=-\frac{\xi_{1}+\xi_{0}+\delta_{\mathrm{d,\pm}}}{2}=\mp\sqrt{\xi_{1}\xi_{0}}.\label{eq:h+-2}
\end{equation}
The expression of Equation (\ref{eq:SpecValDeg}) is real-valued if
and only if $\xi_{1}\xi_{0}>0$, or equivalently

\begin{equation}
\mathrm{\frac{\xi_{0}}{\left|\xi_{0}\right|}=\frac{\xi_{1}}{\left|\xi_{1}\right|}=\pm1.}\label{eq:SignCond}
\end{equation}
Equation (\ref{eq:SignCond}) implies that the equality of resonance
frequencies sign, $\mathrm{sign}\left(\xi_{0}\right)=\mathrm{sign}\left(\xi_{1}\right)$,
is a necessary condition for the eigenvalue degeneracy condition $\Delta_{\mathrm{r},h}=0$
provided that $\delta_{\mathrm{d,\pm}}$ has to be real-valued. From
Equations (\ref{eq:SpecValDeg}) and (\ref{eq:SignCond}), it follows
that the special values of degeneracy $\delta_{\mathrm{d,\pm}}$ can
be expressed as $\delta_{\mathrm{d,\pm}}=-\left(\sqrt{\xi_{1}}\mp\sqrt{\xi_{0}}\right)^{2}$.
\begin{thm}[Nontrivial Jordan canonical form of the companion matrix\label{thm:Jordan}]
 Let's assume that $s_{0}$ is an eigenvalue of the companion matrix
$\mathscr{C}_{\mathrm{r}}$ given in Equation (\ref{eq:MatDiffEq4})
such that its algebraic multiplicity $m\left(s_{0}\right)$$=2$.
Then (i) $s_{0}\neq0$; (ii) $s_{0}$ is either purely real or purely
imaginary; (iii) $-s_{0}$ is also an eigenvalue of the companion
matrix $\mathscr{C}_{\mathrm{r}}$; (iv) $m\left(s_{0}\right)$$=m\left(-s_{0}\right)=2$;
and (v) the nontrivial Jordan canonical form of the companion matrix
$\mathscr{C}_{\mathrm{r}}$ is expressed as
\begin{equation}
\mathscr{\mathscr{J}}_{\mathrm{r}}=\left[\begin{array}{rrrr}
s_{0} & 1 & 0 & 0\\
0 & s_{0} & 0 & 0\\
0 & 0 & -s_{0} & 1\\
0 & 0 & 0 & -s_{0}
\end{array}\right].\label{eq:msJ1a}
\end{equation}
It follows that the eigenvalue degeneracy for the companion matrix
$\mathscr{C}_{\mathrm{r}}$ implies that its Jordan canonical form
$\mathscr{\mathscr{J}}_{\mathrm{r}}$ has two Jordan blocks of size
2. A proof of this theorem can be found in \cite{figotin2020synthesis}.
\end{thm}

\subsection{Eigenvectors and the Jordan basis\label{subsec:Recip-Eigenvectors}}

Theorem \ref{thm:Jordan} states that the degeneracy of eigenvalues
in the companion matrix $\mathscr{C}_{\mathrm{r}}$ given by Equation
(\ref{eq:MatDiffEq4}) implies that its Jordan canonical form consists
of two Jordan blocks as in Equation (\ref{eq:msJ1a}). The Jordan
canonical form corresponding to the companion matrix $\mathscr{C}_{\mathrm{r}}$
in the non-degenerate form, $\Delta_{\mathrm{r},h}\neq0$, is expressed
as

\begin{equation}
\mathscr{J}_{\mathrm{r}}=\left[\begin{array}{rrrr}
\mathrm{i}s_{1} & 0 & 0 & 0\\
0 & -\mathrm{i}s_{1} & 0 & 0\\
0 & 0 & \mathrm{i}s_{2} & 0\\
0 & 0 & 0 & -\mathrm{i}s_{2}
\end{array}\right],\quad\begin{array}{c}
s_{1}=\sqrt{\frac{\xi_{\mathrm{s}}+\sqrt{\Delta_{\mathrm{r},h}}}{2}}\\
s_{2}=\sqrt{\frac{\xi_{\mathrm{s}}-\sqrt{\Delta_{\mathrm{r},h}}}{2}}
\end{array},\quad\xi_{\mathrm{s}}=\xi_{1}+\xi_{0}+\delta,
\end{equation}
where $\xi_{\mathrm{s}}$ is the sum of three resonance frequencies
square defined in Equation (\ref{eq:RecipPara}), $\Delta_{\mathrm{r},h}$
is the discriminant of the quadratic equation defined after Equation
(\ref{eq:h+-}) and $\pm\mathrm{i}s_{i}$ ($i=1,2$) are the corresponding
eigenvalues. Following this, we write eigenvectors corresponding to
the calculated non-degenerate eigenvalues as follows:

\begin{subequations}
\begin{equation}
\mathscr{V}_{\mathrm{r}}=\left[\mathsf{e}_{\mathrm{r},+s_{1}}|\mathsf{e}_{\mathrm{r},-s_{1}}|\mathsf{e}_{\mathrm{r},+s_{2}}|\mathsf{e}_{\mathrm{r},-s_{2}}\right],
\end{equation}

\begin{equation}
\begin{array}{c}
\mathsf{e}_{\mathrm{r},\pm s_{i}}=\left[\begin{array}{r}
\mp\frac{\left(-s_{i}^{2}\right)^{3/2}+\left(\xi_{0}+\delta\right)\sqrt{-s_{i}^{2}}}{\xi_{1}\xi_{0}}\\
\mp\frac{\left(-s_{i}^{2}\right)^{3/2}+\xi_{\mathrm{s}}\sqrt{-s_{i}^{2}}}{\xi_{1}\xi_{0}}\\
1-\frac{s_{i}^{2}}{\xi_{0}}\\
1
\end{array}\right].\end{array}
\end{equation}
\end{subequations}

Next, we investigate two different cases with two special values of
degeneracy\emph{ $\delta_{\mathrm{d,\pm}}$}, which lead to degeneracy
with purely imaginary or purely real degenerate eigenvalues.

\subsubsection{Degeneracy with purely imaginary eigenvalues (purely real eigenfrequencies)}

In the first case, if we consider $\delta=\delta_{\mathrm{d},+}=-\xi_{1}-\xi_{0}+2\sqrt{\xi_{1}\xi_{0}}$,
the characteristic polynomial is rewritten as

\begin{equation}
\chi_{\mathrm{r},+}\left(s\right)=s^{4}+2s^{2}\sqrt{\xi_{1}\xi_{0}}+\xi_{1}\xi_{0}=\left(s^{2}+\sqrt{\xi_{1}\xi_{0}}\right)^{2}=0.
\end{equation}
Then, the degenerate companion matrix $\mathscr{C}_{\mathrm{r},+}$
for this degenerate case is given by

\begin{equation}
\mathscr{C}_{\mathrm{r},+}=\left[\begin{array}{rrrr}
0 & 0 & 1 & 0\\
0 & 0 & 0 & 1\\
-\delta_{\mathrm{d},+}-\xi_{1} & \delta_{\mathrm{d},+} & 0 & 0\\
\xi_{0} & -\xi_{0} & 0 & 0
\end{array}\right],\label{eq:RecipCompImag}
\end{equation}
and the Jordan canonical form of the degenerate companion matrix $\mathscr{C}_{\mathrm{r},+}$
with purely imaginary eigenvalues (purely real eigenfrequencies) is
expressed as

\begin{equation}
\mathscr{J}_{\mathrm{r},+}=\left[\begin{array}{rrrr}
\mathrm{i}s_{\mathrm{e}} & 1 & 0 & 0\\
0 & \mathrm{i}s_{\mathrm{e}} & 0 & 0\\
0 & 0 & -\mathrm{i}s_{\mathrm{e}} & 1\\
0 & 0 & 0 & -\mathrm{i}s_{\mathrm{e}}
\end{array}\right],\quad s_{\mathrm{e}}=\left(\xi_{\mathrm{e}}\right)^{\frac{1}{2}}=\left(\xi_{0}\xi_{1}\right)^{\frac{1}{4}}.
\end{equation}
As a result, the Jordan basis of the degenerate companion matrix $\mathscr{C}_{\mathrm{r},+}$
is obtained as follows:

\begin{equation}
\mathscr{Z}_{\mathrm{r},+}=\left[\begin{array}{rrrr}
\frac{\xi_{0}-\xi_{\mathrm{e}}}{4\mathrm{i}\sqrt{\xi_{\mathrm{e}}}} & \frac{1}{2} & \frac{-\xi_{0}+\xi_{\mathrm{e}}}{4\mathrm{i}\sqrt{\xi_{\mathrm{e}}}} & \frac{1}{2}\\
\frac{\xi_{0}}{4\mathrm{i}\sqrt{\xi_{\mathrm{e}}}} & 0 & -\frac{\xi_{0}}{4\mathrm{i}\sqrt{\xi_{\mathrm{e}}}} & 0\\
\frac{\xi_{0}\left(\xi_{\mathrm{e}}-\xi_{1}\right)}{4\xi_{\mathrm{e}}} & \frac{\xi_{0}\left(\xi_{\mathrm{e}}-3\xi_{1}\right)}{4\mathrm{i}\xi_{\mathrm{e}}\sqrt{\xi_{\mathrm{e}}}} & \frac{\xi_{0}\left(\xi_{\mathrm{e}}-\xi_{1}\right)}{4\xi_{\mathrm{e}}} & -\frac{\xi_{0}\left(\xi_{\mathrm{e}}-3\xi_{1}\right)}{4\mathrm{i}\xi_{\mathrm{e}}\sqrt{\xi_{\mathrm{e}}}}\\
-\frac{\xi_{0}^{2}\left(2\xi_{1}-\xi_{\mathrm{e}}\right)}{4\xi_{\mathrm{e}}\left(\xi_{0}-2\xi_{\mathrm{e}}\right)} & -\frac{\xi_{0}^{2}\left(2\xi_{1}-\xi_{\mathrm{e}}\right)}{4\mathrm{i}\xi_{\mathrm{e}}\sqrt{\xi_{\mathrm{e}}}\left(\xi_{0}-2\xi_{\mathrm{e}}\right)} & -\frac{\xi_{0}^{2}\left(2\xi_{1}-\xi_{\mathrm{e}}\right)}{4\xi_{\mathrm{e}}\left(\xi_{0}-2\xi_{\mathrm{e}}\right)} & \frac{\xi_{0}^{2}\left(2\xi_{1}-\xi_{\mathrm{e}}\right)}{4\mathrm{i}\xi_{e}\sqrt{\xi_{\mathrm{e}}}\left(\xi_{0}-2\xi_{\mathrm{e}}\right)}
\end{array}\right].
\end{equation}
Note that the columns of matrix $\mathscr{Z}_{\mathrm{r},+}$ form
the Jordan bases of the corresponding degenerate companion matrix
$\mathscr{C}_{\mathrm{r},+}$, and each column of $\mathscr{Z}_{\mathrm{r},+}$
are the generalized eigenvectors of the corresponding eigenvalues.

\subsubsection{Degeneracy with purely real eigenvalues (purely imaginary eigenfrequencies)}

In the second case, if we consider $\delta=\delta_{\mathrm{d},-}=-\xi_{1}-\xi_{0}-2\sqrt{\xi_{1}\xi_{0}}$,
the characteristic polynomial is given by

\begin{equation}
\chi_{\mathrm{r},-}\left(s\right)=s^{4}-2s^{2}\sqrt{\xi_{1}\xi_{0}}+\xi_{1}\xi_{0}=\left(s^{2}-\sqrt{\xi_{1}\xi_{0}}\right)^{2}=0.
\end{equation}
The corresponding degenerate companion matrix $\mathscr{C}_{\mathrm{r},-}$
is expressed by

\begin{equation}
\mathscr{C}_{\mathrm{r,-}}=\left[\begin{array}{rrrr}
0 & 0 & 1 & 0\\
0 & 0 & 0 & 1\\
-\delta_{\mathrm{d},-}-\xi_{1} & \delta_{\mathrm{d},-} & 0 & 0\\
\xi_{0} & -\xi_{0} & 0 & 0
\end{array}\right].\label{eq:RecipCompReal}
\end{equation}
Moreover, the Jordan canonical form of the corresponding degenerate
companion matrix $\mathscr{C}_{\mathrm{r},-}$ in Equation (\ref{eq:RecipCompReal})
with purely real eigenvalues (purely imaginary eigenfrequencies) is
written as

\begin{equation}
\mathscr{J}_{\mathrm{r,-}}=\left[\begin{array}{rrrr}
s_{\mathrm{e}} & 1 & 0 & 0\\
0 & s_{\mathrm{e}} & 0 & 0\\
0 & 0 & -s_{\mathrm{e}} & 1\\
0 & 0 & 0 & -s_{\mathrm{e}}
\end{array}\right].
\end{equation}
Then, the Jordan basis of the degenerate companion matrix $\mathscr{C}_{\mathrm{r,-}}$
is expressed as

\begin{equation}
\mathscr{Z}_{\mathrm{r,-}}=\left[\begin{array}{rrrr}
\frac{\xi_{0}+\xi_{\mathrm{e}}}{4\sqrt{\xi_{\mathrm{e}}}} & \frac{1}{2} & -\frac{\xi_{0}+\xi_{\mathrm{e}}}{4\sqrt{\xi_{\mathrm{e}}}} & \frac{1}{2}\\
\frac{\xi_{0}}{4\sqrt{\xi_{\mathrm{e}}}} & 0 & -\frac{\xi_{0}}{4\sqrt{\xi_{\mathrm{e}}}} & 0\\
\frac{\xi_{0}\left(\xi_{\mathrm{e}}+\xi_{1}\right)}{4\xi_{\mathrm{e}}} & \frac{\xi_{0}\left(\xi_{\mathrm{e}}+3\xi_{1}\right)}{4\xi_{\mathrm{e}}\sqrt{\xi_{\mathrm{e}}}} & \frac{\xi_{0}\left(\xi_{\mathrm{e}}+\xi_{1}\right)}{4\xi_{\mathrm{e}}} & -\frac{\xi_{0}\left(\xi_{\mathrm{e}}+3\xi_{1}\right)}{4\xi_{\mathrm{e}}\sqrt{\xi_{\mathrm{e}}}}\\
\frac{\xi_{0}^{2}\left(2\xi_{1}+\xi_{\mathrm{e}}\right)}{4\xi_{\mathrm{e}}\left(\xi_{0}+2\xi_{\mathrm{e}}\right)} & \frac{\xi_{0}^{2}\left(2\xi_{1}+\xi_{\mathrm{e}}\right)}{4\xi_{\mathrm{e}}\sqrt{\xi_{\mathrm{e}}}\left(\xi_{0}+2\xi_{\mathrm{e}}\right)} & \frac{\xi_{0}^{2}\left(2\xi_{1}+\xi_{\mathrm{e}}\right)}{4\xi_{\mathrm{e}}\left(\xi_{0}+2\xi_{\mathrm{e}}\right)} & -\frac{\xi_{0}^{2}\left(2\xi_{1}+\xi_{\mathrm{e}}\right)}{4\xi_{\mathrm{e}}\sqrt{\xi_{\mathrm{e}}}\left(\xi_{0}+2\xi_{\mathrm{e}}\right)}
\end{array}\right].
\end{equation}

\section{Lagrangian and Hamiltonian Structures and Their Relation\label{sec:Lagrangian-and-Hamiltonian}}

In this section we provide a general overview of the Lagrangian and
Hamiltonian structures. We explain the relationship between these
two structures in detail. Then, we demonstrate the Lagrangian and
Hamiltonian structures for the PRC. Ultimately, readers will gain
a comprehensive understanding of these mathematical frameworks and
their applicability to studying systems like the PRC.

\subsection{Lagrangian structure\label{subsec:RecipLag}}

The Lagrangian for a linear system in the general form is a quadratic
function (bilinear form) of the circuit state vector $\mathrm{\check{q}}$
(column vector that contains charges) and its time derivatives $\partial_{t}\mathrm{\check{q}}$,
that is

\begin{gather}
\mathcal{L}=\mathcal{L}\left(\mathrm{\check{q}},\partial_{t}\mathrm{\check{q}}\right)=\frac{1}{2}\left[\begin{array}{r}
\mathrm{\check{q}}\\
\partial_{t}\mathrm{\check{q}}
\end{array}\right]^{\mathrm{T}}\mathrm{M}_{\mathrm{L}}\left[\begin{array}{r}
\mathrm{\check{q}}\\
\partial_{t}\mathrm{\check{q}}
\end{array}\right],\quad\mathrm{M}_{\mathrm{L}}=\left[\begin{array}{rr}
-\eta & \theta^{\mathrm{T}}\\
\theta & \alpha
\end{array}\right],\label{eq:Lag1}
\end{gather}
where $\alpha,\eta$ and $\theta$ are $2\times2$ matrices (for our
two-loops circuit) with real-valued entries. Moreover, we assume symmetric
matrices, that is $\alpha=\alpha^{\mathrm{T}}$ and $\eta=\eta^{\mathrm{T}}$.
Accordingly, we have the Lagrangian,
\begin{equation}
\mathcal{L}=\frac{1}{2}\partial_{t}\mathrm{\check{q}}^{\mathrm{T}}\alpha\partial_{t}\mathrm{\check{q}}+\partial_{t}\mathrm{\check{q}}^{\mathrm{T}}\theta\mathrm{\check{q}}-\frac{1}{2}\mathrm{\check{q}}^{\mathrm{T}}\eta\mathrm{\check{q}}.\label{eq:Lag2}
\end{equation}
As a result of Hamilton's principle, the system evolution can be explained
by the Euler-Lagrange equations as
\begin{equation}
\frac{d}{dt}\left(\frac{\partial\mathcal{L}}{\partial\dot{\mathrm{\check{q}}}}\right)-\frac{\partial\mathcal{L}}{\partial\mathrm{\check{q}}}=0,\label{eq:Lag3}
\end{equation}
which, based on Equation (\ref{eq:Lag2}), it can be written in the
following form of a second-order vector ordinary differential equation,
\begin{equation}
\alpha\partial_{t}^{2}\mathrm{\check{q}}+\left(\theta-\theta^{\mathrm{T}}\right)\partial_{t}\mathrm{\check{q}}+\eta\mathrm{\check{q}}=0.\label{eq:Lag4}
\end{equation}
It is notable that matrix $\theta$ appears in Equation (\ref{eq:Lag4})
through its skew-symmetric component $\left(\theta-\theta^{\mathrm{T}}\right)/2$
justifying as a possibility to impose the skew-symmetry assumption
on $\theta$, that is $\theta^{\mathrm{T}}=-\theta$. Then, under
the assumption of $\theta^{\mathrm{T}}=-\theta$, Equation (\ref{eq:Lag4})
is rewritten with the skew-symmetric $\theta$ as
\begin{equation}
\alpha\partial_{t}^{2}\mathrm{\check{q}}+2\theta\partial_{t}\mathrm{\check{q}}+\eta\mathrm{\check{q}}=0.\label{eq:Lag5}
\end{equation}
For the PRC that we introduced and analyzed in the previous sections,
Equation (\ref{eq:Lag5}) and the required coefficients by considering
$\mathrm{\check{q}}=\mathrm{q}_{\mathrm{r}}=\left[q_{1},q_{0}\right]^{\mathrm{T}}$
are rewritten as

\begin{subequations}
\begin{equation}
\alpha_{\mathrm{r}}\partial_{t}^{2}\mathrm{q}_{\mathrm{r}}+2\theta_{\mathrm{r}}\partial_{t}\mathrm{q}_{\mathrm{r}}+\eta_{\mathrm{r}}\mathrm{q}_{\mathrm{r}}=0,\label{eq:Lag6}
\end{equation}

\begin{equation}
\alpha_{\mathrm{r}}=\left[\begin{array}{cc}
L_{1} & 0\\
0 & L_{0}
\end{array}\right],\quad\theta_{\mathrm{r}}=\left[\begin{array}{cc}
0 & 0\\
0 & 0
\end{array}\right],\quad\eta_{\mathrm{r}}=\left[\begin{array}{rr}
\frac{1}{C_{1}}+\frac{1}{C_{0}} & -\frac{1}{C_{0}}\\
-\frac{1}{C_{0}} & \frac{1}{C_{0}}
\end{array}\right].\label{eq:Lag_alptheeta}
\end{equation}
\end{subequations}
We write the Lagrangian equation in the below form:

\begin{subequations}
\begin{equation}
\mathcal{L}_{\mathrm{r}}\left(\mathrm{q}_{\mathrm{r}},\partial_{t}\mathrm{q}_{\mathrm{r}}\right)=\frac{1}{2}\left[\begin{array}{r}
\mathrm{q}_{\mathrm{r}}\\
\partial_{t}\mathrm{q}_{\mathrm{r}}
\end{array}\right]^{\mathrm{T}}\mathrm{M}_{\mathrm{L,r}}\left[\begin{array}{r}
\mathrm{q}_{\mathrm{r}}\\
\partial_{t}\mathrm{q}_{\mathrm{r}}
\end{array}\right],
\end{equation}

\begin{equation}
\mathrm{M}_{\mathrm{L,r}}=\left[\begin{array}{rrrr}
-\frac{1}{C_{1}}-\frac{1}{C_{0}} & \frac{1}{C_{0}} & 0 & 0\\
\frac{1}{C_{0}} & -\frac{1}{C_{0}} & 0 & 0\\
0 & 0 & L_{1} & 0\\
0 & 0 & 0 & L_{0}
\end{array}\right].
\end{equation}
\end{subequations}
It follows from Equation (\ref{eq:Lag4}) that the necessary and sufficient
condition for nonreciprocity is $\theta\neq\theta^{\mathrm{T}}$.
Indeed, if $\theta=\theta^{\mathrm{T}}$, the circuit will not show
nonreciprocity properties. Because in the case of $\theta=\theta^{\mathrm{T}}$,
the second term with $\partial_{t}\mathrm{q}$ disappears from Equation
(\ref{eq:Lag4}) and all frequencies related to Equation (\ref{eq:Lag4})
will be the same as in the case of zero $\theta$ which leads to time
symmetry.

\subsection{Hamiltonian\label{subsec:Ham}}

Alternatively, we can use the Hamilton equations associated with the
Hamiltonian $\mathcal{H}$ instead of the second-order vector ordinary
differential equations of Equation (\ref{eq:Lag4}). Let us suppose
that the system is described by a time-dependent column vector $\mathrm{\check{q}}$
and its dynamic is governed by a Hamiltonian $\mathcal{H}=\mathcal{H}\left(\mathrm{\check{p}},\mathrm{\check{q}}\right)$,
where $\mathrm{\check{p}}$ is the column vector system momentum.
We introduce the Hamiltonian representation in order to present the
system information in compact matrix form,
\begin{gather}
\mathcal{H}=\mathcal{H}\left(\mathrm{\check{p}},\mathrm{\check{q}}\right)=\mathrm{\check{p}}^{\mathrm{T}}\partial_{t}\mathrm{\check{q}}-\mathcal{L}\left(\mathrm{\check{q}},\partial_{t}\mathrm{\check{q}}\right),\label{eq:Hamil1a}
\end{gather}
where the system momentum $\mathrm{\check{p}}$ is given by

\begin{gather}
\mathrm{\check{p}}=\frac{\partial\mathcal{L}}{\partial\dot{\mathrm{\check{q}}}}=\alpha\partial_{t}\mathrm{\check{q}}+\theta\mathrm{\check{q}}.\label{eq:Hamil1b}
\end{gather}
According to Equation (\ref{eq:Hamil1b}), current $\partial_{t}\mathrm{\check{q}}$
(i.e., velocity of charges) and momentum $\mathrm{\check{p}}$ vectors
are related as follows:
\begin{equation}
\partial_{t}\mathrm{\check{q}}=\alpha^{-1}\left(\mathrm{\check{p}}-\theta\mathrm{\check{q}}\right).\label{eq:Hamil2}
\end{equation}
Consequently, the Hamiltonian $\mathcal{H}$ expressed in Equation
(\ref{eq:Hamil1a}) is given by
\begin{gather}
\mathcal{H}\left(\mathrm{\check{p}},\mathrm{\check{q}}\right)=\frac{1}{2}\left[\left(\mathrm{\check{p}}-\theta\mathrm{\check{q}}\right)^{\mathrm{T}}\alpha^{-1}\left(\mathrm{\check{p}}-\theta\mathrm{\check{q}}\right)+\mathrm{\check{q}}^{\mathrm{T}}\eta\mathrm{\check{q}}\right]=\frac{1}{2}\partial_{t}\mathrm{\check{q}}^{\mathrm{T}}\alpha\partial_{t}\mathrm{\check{q}}+\frac{1}{2}\mathrm{\check{q}}^{\mathrm{T}}\eta\mathrm{\check{q}}.\label{eq:Hamil3}
\end{gather}
Also, Equations (\ref{eq:Hamil1a}), (\ref{eq:Hamil1b}) and (\ref{eq:Hamil2})
imply

\begin{gather}
\left[\begin{array}{r}
\mathrm{\check{q}}\\
\mathrm{\check{p}}
\end{array}\right]=\left[\begin{array}{lr}
\mathbb{I}_{2} & 0_{2}\\
\theta & \alpha
\end{array}\right]\left[\begin{array}{r}
\mathrm{\check{q}}\\
\partial_{t}\mathrm{\check{q}}
\end{array}\right],\quad\left[\begin{array}{r}
\mathrm{\check{q}}\\
\partial_{t}\mathrm{\check{q}}
\end{array}\right]=\left[\begin{array}{rr}
\mathbb{I}_{2} & 0_{2}\\
-\alpha^{-1} & \theta\alpha^{-1}
\end{array}\right]\left[\begin{array}{l}
\mathrm{\check{q}}\\
\mathrm{\check{p}}
\end{array}\right].\label{eq:Hamil4}
\end{gather}
where $\mathbb{I}_{2}$ is the $2\times2$ identity matrix and $0_{2}$
is the $2\times2$ zero matrix. We also know that Hamiltonian $\mathcal{H}$
can be interpreted as the energy of the system that is a conserved
quantity, so $\partial_{t}\mathcal{H}\left(\mathrm{\check{p}},\mathrm{\check{q}}\right)=0$.
The function $\mathcal{H}\left(\mathrm{\check{p}},\mathrm{\check{q}}\right)$
defined by Equation (\ref{eq:Hamil3}) can be recast into the following
form of Hamiltonian
\begin{equation}
\mathcal{H}\left(\mathrm{\check{p}},\mathrm{\check{q}}\right)=\frac{1}{2}\left[\begin{array}{r}
\mathrm{\check{q}}\\
\mathrm{\check{p}}
\end{array}\right]^{\mathrm{T}}\mathrm{M}_{\mathrm{H}}\left[\begin{array}{r}
\mathrm{\check{q}}\\
\mathrm{\check{p}}
\end{array}\right],\label{eq:Hamil5}
\end{equation}
where the matrix $\mathrm{M}_{\mathrm{H}}$ is $4\times4$ with the
below block form
\begin{gather}
\mathrm{M}_{\mathrm{H}}=\left[\begin{array}{rr}
\theta^{\mathrm{T}}\alpha^{-1}\theta+\eta & -\theta^{\mathrm{T}}\alpha^{-1}\\
-\alpha^{-1}\theta & \alpha^{-1}
\end{array}\right]=\left[\begin{array}{rr}
\mathbb{I}_{2} & -\theta^{\mathrm{T}}\\
0_{2} & \mathbb{I}_{2}
\end{array}\right]\left[\begin{array}{rr}
\eta & 0_{2}\\
0_{2} & \alpha^{-1}
\end{array}\right]\left[\begin{array}{rr}
\mathbb{I}_{2} & 0_{2}\\
-\theta & \mathbb{I}_{2}
\end{array}\right],\label{eq:Hamil6}
\end{gather}
and the required parameters for the PRC are defined in Equation (\ref{eq:Lag_alptheeta}).
Then, the matrix $\mathrm{M}_{\mathrm{H}}$ for the PRC is expressed
as

\begin{gather}
\mathrm{M}_{\mathrm{H,r}}=\left[\begin{array}{rrrr}
\frac{1}{C_{1}}+\frac{1}{C_{0}} & -\frac{1}{C_{0}} & 0 & 0\\
-\frac{1}{C_{0}} & \frac{1}{C_{0}} & 0 & 0\\
0 & 0 & \frac{1}{L_{1}} & 0\\
0 & 0 & 0 & \frac{1}{L_{0}}
\end{array}\right].\label{eq:Hamil7}
\end{gather}
Also, the system momentum in Equation (\ref{eq:Hamil1b}) by considering
$\mathrm{\check{q}}=\mathrm{q}_{\mathrm{r}}=\left[q_{1},q_{0}\right]^{\mathrm{T}}$
for the PRC is expressed by

\begin{equation}
\mathrm{p}_{\mathrm{r}}=\frac{\partial\mathcal{L}_{\mathrm{r}}}{\partial\dot{\mathrm{q}}_{\mathrm{r}}}=\alpha_{\mathrm{r}}\partial_{t}\mathrm{q}_{\mathrm{r}}+\theta_{\mathrm{r}}\mathrm{q}_{\mathrm{r}}=\partial_{t}\mathrm{q}_{\mathrm{r}}\left[\begin{array}{cc}
L_{1} & 0\\
0 & L_{0}
\end{array}\right].\label{eq:Hamil2Recip}
\end{equation}
As we observe in Equation (\ref{eq:Hamil2Recip}), $\theta_{\mathrm{r}}=0$
for the reciprocal case, which demonstrates the relation between the
\textit{momentum} $\mathrm{p}_{\mathrm{r}}$ and the \textit{current}
$\partial_{t}\mathrm{q}_{\mathrm{r}}$ (i.e., velocity of charges)
does not depend on $\mathrm{q}_{\mathrm{r}}$ and depends only on
$\partial_{t}\mathrm{q}_{\mathrm{r}}$. The Hamiltonian form of the
Euler-Lagrange formulation of Equation (\ref{eq:Lag3}) is given by

\begin{gather}
\partial_{t}\left[\begin{array}{r}
\mathrm{\check{q}}\\
\mathrm{\check{p}}
\end{array}\right]=\mathbb{J}\mathrm{M}_{\mathrm{H}}\left[\begin{array}{r}
\mathrm{\check{q}}\\
\mathrm{\check{p}}
\end{array}\right],\quad\mathbb{J}=\left[\begin{array}{rr}
0_{2} & \mathbb{I}_{2}\\
-\mathbb{I}_{2} & 0_{2}
\end{array}\right],\label{eq:Hamil8}
\end{gather}
where the matrix $\mathbb{J}$ has the following properties
\begin{equation}
\mathbb{J}=-\mathbb{J}^{\mathrm{T}}=-\mathbb{J}^{-1}.\label{eq:Hamil9}
\end{equation}
Based on Equations (\ref{eq:Hamil6}) and (\ref{eq:Hamil8}), we write
\begin{gather}
\mathbb{J}\mathrm{M}_{\mathrm{H}}=\left[\begin{array}{rr}
-\alpha^{-1}\theta & \alpha^{-1}\\
-\theta^{\mathrm{T}}\alpha^{-1}\theta-\eta & \theta^{\mathrm{T}}\alpha^{-1}
\end{array}\right].\label{eq:Hamil10}
\end{gather}
Ultimately, the Hamiltonian matrix for the PRC is expressed as
\begin{equation}
\mathscr{H}_{\mathrm{r}}=\mathbb{J}\mathrm{M}_{\mathrm{H,r}}=\left[\begin{array}{rrrr}
0 & 0 & \frac{1}{L_{1}} & 0\\
0 & 0 & 0 & \frac{1}{L_{0}}\\
-\frac{1}{C_{1}}-\frac{1}{C_{0}} & \frac{1}{C_{0}} & 0 & 0\\
\frac{1}{C_{0}} & -\frac{1}{C_{0}} & 0 & 0
\end{array}\right].\label{eq:Hamil11}
\end{equation}

According to Equation (\ref{eq:Hamil6}), we know $\mathrm{M}_{\mathrm{H}}^{\mathrm{T}}=\mathrm{M}_{\mathrm{H}}$
and
\begin{gather}
\mathscr{H}^{\mathrm{T}}=-\mathrm{M}_{\mathrm{H}}\mathbb{J}=-\mathbb{J}\left[-\mathbb{J}\mathrm{M}_{\mathrm{H}}\right]\mathbb{J}=\mathbb{J}^{-1}\left[-\mathscr{H}\right]\mathbb{J},\label{eq:Hamil12}
\end{gather}
where demonstrate that the transpose of the matrix $\mathscr{H}$
is similar to $-\mathscr{H}$.

\subsection{Relationship between the Lagrangian and Hamiltonian\label{subsec:Lag-to-Ham}}

By considering the assumption that $\alpha$ is invertible matrix,
according to Equations (\ref{eq:Lag1}), (\ref{eq:Hamil5}) and (\ref{eq:Hamil6}),
we have
\begin{gather}
\mathrm{M}_{\mathrm{L}}=\left[\begin{array}{rr}
-\eta & \theta^{\mathrm{T}}\\
\theta & \alpha
\end{array}\right],\quad\mathrm{M}_{\mathrm{H}}=\left[\begin{array}{rr}
\beta & \gamma^{\mathrm{T}}\\
\gamma & \rho
\end{array}\right]=\left[\begin{array}{rr}
\theta^{\mathrm{T}}\alpha^{-1}\theta+\eta & -\theta^{\mathrm{T}}\alpha^{-1}\\
-\alpha^{-1}\theta & \alpha^{-1}
\end{array}\right],\label{eq:LaH1a}
\end{gather}
implying that
\begin{gather}
\rho=\alpha^{-1},\quad\gamma=-\alpha^{-1}\theta,\quad\beta=\theta^{\mathrm{T}}\alpha^{-1}\theta+\eta,\label{eq:LaH1b}
\end{gather}
or
\begin{gather}
\alpha=\rho^{-1},\quad\theta=-\rho^{-1}\gamma,\quad\eta=\beta-\gamma^{\mathrm{T}}\rho^{-1}\gamma.\label{eq:LaH1c}
\end{gather}

By using the transformations between coefficients described above,
the coefficients of the matrix $\mathrm{M}_{\mathrm{L}}$, i.e, $\eta$,
$\theta$ and $\alpha$, can be converted into the coefficients of
the matrix $\mathrm{M}_{\mathrm{H}}$, i.e., $\beta$, $\gamma$ and
$\rho$, and vice versa. As a result, we can construct the Hamiltonian
representation from the Lagrangian representation and vice versa.

\section{Gyrator-based Nonreciprocal Circuit (GNC)\label{sec:Gyrator-based-Circuit}}

In this section, we summarize the implementation of a gyrator-based
circuit to obtain the nontrivial Jordan canonical form that was already
proposed and analyzed in our previous works \cite{figotin2020synthesis,figotin2021perturbations,nikzamir2021demonstration,rouhi2022exceptional,rouhi2022high}.
The gyrator as a basic circuit element was initially introduced by
Tellegen \cite[Chapter 18]{tellegen1948gyrator,helszajn1992microwave}.
In electric circuits without any nonreciprocal element such as a gyrator
the impedance matrix of $n$-port system $\mathrm{Z}_{n\times n}$
is always symmetric, i.e., $\mathrm{Z}=\mathrm{Z}^{\mathrm{T}}$.
In contrast, the presence of gyrators \textit{may} result in an asymmetric
impedance matrix, i.e, $\mathrm{Z}\neq\mathrm{Z}^{\mathrm{T}}$, which
can be interpreted as the nonreciprocity property. The ideal gyrator
is characterized as a lossless two-port circuit element with the below
relationship between the input and output voltages ($v_{1}$, $v_{2}$)
and currents ($i_{1}$, $i_{2}$) (see Figure \ref{fig:GyratorCircuit}):

\begin{figure}
\begin{centering}
\includegraphics[width=0.41\textwidth]{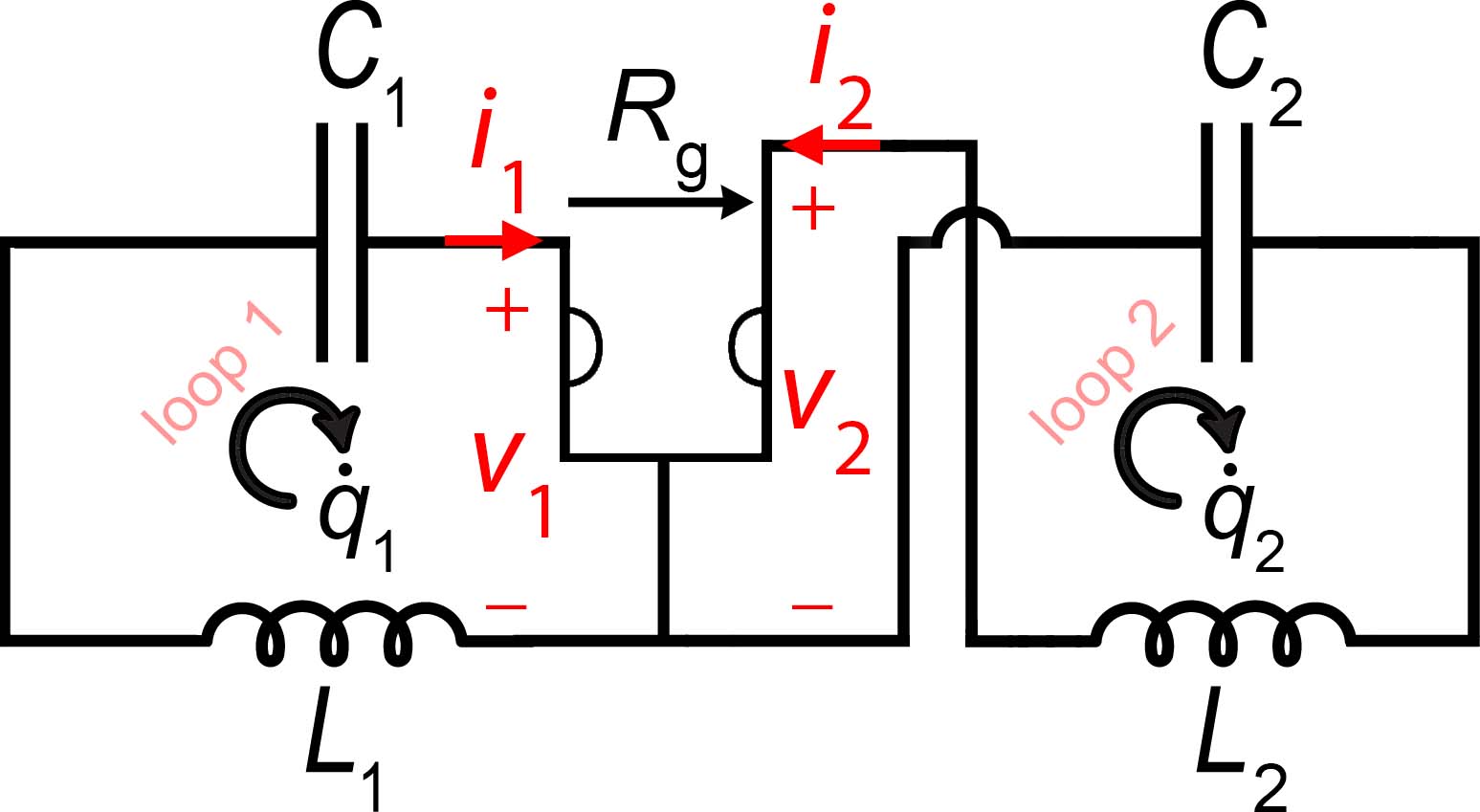}
\par\end{centering}
\caption{The two-loop nonreciprocal circuit coupled by an ideal gyrator. For
specific combinations of values $L_{1}$, $C_{1}$, $L_{2}$, and
$C_{2}$, the evolution matrix of the nonreciprocal circuit develops
a degeneracy, and its nontrivial Jordan canonical form consists of
exactly two Jordan blocks of size 2. A change in the sign of $R_{\mathrm{g}}$
is equivalent to reversing the gyration resistance direction (indicated
by the arrow below $R_{\mathrm{g}}$) in this circuit.\label{fig:GyratorCircuit}}
\end{figure}
\begin{subequations}
\begin{equation}
v_{1}=-R_{\mathrm{g}}i_{2}=-R_{\mathrm{g}}\dot{q}_{2},
\end{equation}

\begin{equation}
v_{2}=R_{\mathrm{g}}i_{1}=R_{\mathrm{g}}\dot{q}_{1}.
\end{equation}
\end{subequations}
where $q_{j}$ ($j=1,2$) are the charges associated with currents
of gyrator. The coefficient $R_{\mathrm{g}}$ is called \textit{gyration
resistance} and the corresponding antisymmetric impedance matrix of
the gyrator is given by

\begin{equation}
Z_{\mathrm{g}}=\left[\begin{array}{rr}
0 & R_{\mathrm{g}}\\
-R_{\mathrm{g}} & 0
\end{array}\right]=\left(-Z_{\mathrm{g}}\right)^{\mathrm{T}}.
\end{equation}
The Lagrangian associated with the gyrator element is given by

\begin{equation}
\mathcal{L}_{\mathrm{g}}=\frac{R_{\mathrm{g}}\left(q_{1}\partial_{t}q_{2}-q_{2}\partial_{t}q_{1}\right)}{2}.
\end{equation}

The simple form of the GNC to obtain second-order degeneracy is shown
in Figure \ref{fig:GyratorCircuit} and analyzed in the following.
The circuit is composed of two LC loops coupled by a gyrator, and
we study this circuit without imposing any assumptions on the circuit
parameters $L_{1}$, $C_{1}$, $L_{2}$, $C_{2}$ and $R_{\mathrm{g}}$
except that they are all real and non-zero. The Lagrangian of the
GNC is described as

\begin{subequations}
\begin{equation}
\left.\mathcal{L}_{\mathrm{nr}}\left(\mathrm{q}_{\mathrm{nr}},\partial_{t}\mathrm{q}_{\mathrm{nr}}\right)\right|_{\Theta_{\mathrm{nr}}}=\frac{L_{1}\left(\partial_{t}q_{1}\right)^{2}}{2}+\frac{L_{2}\left(\partial_{t}q_{2}\right)^{2}}{2}-\frac{\left(q_{1}\right)^{2}}{2C_{1}}-\frac{\left(q_{2}\right)^{2}}{2C_{2}}+\frac{R_{\mathrm{g}}\left(q_{1}\partial_{t}q_{2}-q_{2}\partial_{t}q_{1}\right)}{2},\label{eq:GyrLag}
\end{equation}

\begin{equation}
q_{j}\left(t\right)=\int^{t}i_{j}\left(t'\right)\,\mathrm{d}t',\quad j=1,2,
\end{equation}
\end{subequations}
where the last term in Equation (\ref{eq:GyrLag}) is the source of
nonreciprocity in the circuit and we will discuss it later in Subsection
\ref{subsec:ReciprocityLageq}. Also, $q_{j}$ and $i_{j}$ are the
charges and the currents associated with loops of the GNC depicted
in Figure \ref{fig:GyratorCircuit}, $\Theta_{\mathrm{nr}}=\left\{ L_{1},L_{2},C_{1},C_{2},R_{\mathrm{g}}\right\} $
is the set of circuit's parameters and we define the vector of charges
for the GNC as $\mathrm{q}_{\mathrm{nr}}=\left[q_{1},q_{2}\right]^{\mathrm{T}}$.
In nonreciprocal systems, due to the breaking of time reversal symmetry,
the Lagrangian associated with the GNC leads to $\mathcal{L}_{\mathrm{nr}}\left(\mathrm{q}_{\mathrm{nr}},\partial_{t}\mathrm{q}_{\mathrm{nr}}\right)\neq\mathcal{L}_{\mathrm{nr}}\left(\mathrm{q}_{\mathrm{nr}},-\partial_{t}\mathrm{q}_{\mathrm{nr}}\right)$.
The corresponding evolution equations after simplification, i.e.,
the Euler-Lagrange equations, are expressed as
\begin{subequations}
\begin{gather}
L_{1}\ddot{q}_{1}-R_{\mathrm{g}}\dot{q_{2}}+\frac{1}{C_{1}}q_{1}=0,\label{eq:GyrEL1}
\end{gather}
\begin{gather}
L_{2}\ddot{q}_{2}+R_{\mathrm{g}}\dot{q_{1}}+\frac{1}{C_{2}}q_{2}=0.\label{eq:GyrEL2}
\end{gather}
\end{subequations}

The second-order differential equations in Equations (\ref{eq:GyrEL1})
and (\ref{eq:GyrEL2}) are reduced to the standard first-order vector
differential equation as
\begin{equation}
\partial_{t}\mathsf{q}_{\mathrm{nr}}=\mathscr{C}_{\mathrm{nr}}\mathsf{q}_{\mathrm{nr}},\quad\mathscr{C}_{\mathrm{nr}}=\left[\begin{array}{rrrr}
0 & 0 & 1 & 0\\
0 & 0 & 0 & 1\\
-\frac{1}{L_{1}C_{1}} & 0 & 0 & \frac{R_{g}}{L_{1}}\\
0 & -\frac{1}{L_{2}C_{2}} & -\frac{R_{g}}{L_{2}} & 0
\end{array}\right],\label{eq:GyrCirMat}
\end{equation}
where the circuit state vector is defined as $\mathsf{q}_{\mathrm{nr}}=\left[\mathrm{q}_{\mathrm{nr}},\partial_{t}\mathrm{q}_{\mathrm{nr}}\right]^{\mathrm{T}}$and
$\mathscr{C}_{\mathrm{nr}}$ is the $4\times4$ circuit matrix for
the GNC. A matrix representation of evolution equations is easily
achieved by recasting them as follows:

\begin{equation}
\mathrm{A}_{\mathrm{nr}}\left(s\right)\mathrm{q}_{\mathrm{nr}}=0,\quad\mathrm{A}_{\mathrm{nr}}\left(s\right)=\left[\begin{array}{rr}
s^{2}+\frac{1}{L_{1}C_{1}} & -s\frac{R_{\mathrm{g}}}{L_{1}}\\
s\frac{R_{\mathrm{g}}}{L_{2}} & s^{2}+\frac{1}{L_{2}C_{2}}
\end{array}\right],\quad\mathrm{q}_{\mathrm{nr}}=\left[\begin{array}{c}
q_{1}\\
q_{2}
\end{array}\right],\label{eq:GyrMatPol1}
\end{equation}
where $\mathrm{A}_{\mathrm{nr}}\left(s\right)$ is a $2\times2$ monic
matrix polynomial of $s$ of the degree 2 and it is rewritten as

\begin{subequations}
\begin{equation}
\mathrm{A}_{\mathrm{nr}}\left(s\right)=\sum_{j=0}^{2}s^{j}\mathrm{A}_{\mathrm{nr},j}=s^{2}\mathrm{A}_{\mathrm{nr},2}+s\mathrm{A}_{\mathrm{nr},1}+\mathrm{A}_{\mathrm{nr},0},\label{eq:GyrMatPol2}
\end{equation}

\begin{equation}
\mathrm{A}_{\mathrm{nr},2}=\left[\begin{array}{rr}
1 & 0\\
0 & 1
\end{array}\right],\quad\mathrm{A}_{\mathrm{nr},1}=\left[\begin{array}{rr}
0 & -\frac{R_{\mathrm{g}}}{L_{1}}\\
\frac{R_{\mathrm{g}}}{L_{2}} & 0
\end{array}\right],\quad\mathrm{A}_{\mathrm{nr},0}=\left[\begin{array}{rr}
\frac{1}{L_{1}C_{1}} & 0\\
0 & \frac{1}{L_{2}C_{2}}
\end{array}\right].
\end{equation}
\end{subequations}

The eigenvalue problem corresponding to the Equation (\ref{eq:GyrCirMat})
is
\begin{equation}
\left(s\mathbb{I}_{4}-\mathscr{C}_{\mathrm{nr}}\right)\mathsf{q}_{\mathrm{nr,s}}=0,\quad\mathsf{q}_{\mathrm{nr,s}}=\left[\begin{array}{r}
\mathrm{q}_{\mathrm{nr}}\\
s\mathrm{q}_{\mathrm{nr}}
\end{array}\right].\label{eq:GyrCharPol1}
\end{equation}
The associated characteristic polynomial for the GNC is expressed
as
\begin{equation}
\chi_{\mathrm{nr}}\left(s\right)=\det\left\{ A_{\mathrm{nr}}\left(s\right)\right\} =\det\left\{ s\mathbb{I}_{4}-\mathscr{C}_{\mathrm{nr}}\right\} =s^{4}+s^{2}\left(\frac{1}{L_{1}C_{1}}+\frac{1}{L_{2}C_{2}}+\frac{R_{\mathrm{g}}^{2}}{L_{1}L_{2}}\right)+\frac{1}{L_{1}C_{1}L_{2}C_{2}},\label{eq:GyrCharPol2}
\end{equation}
and the eigenvalues of the eigenvalue problem can be calculated from
the characteristic polynomial $\chi_{\mathrm{nr}}\left(s\right)=0$.
The further analytical developments suggest to introduce the following
variables for the circuit
\begin{equation}
\xi_{1}=\frac{1}{L_{1}C_{1}},\quad\xi_{2}=\frac{1}{L_{2}C_{2}}.\label{eq:GyrCharPol3}
\end{equation}
In Equation (\ref{eq:GyrCharPol3}), $\sqrt{\xi_{1}}$ is the resonance
frequency of the first loop and $\sqrt{\xi_{2}}$ is the resonance
frequency of the second loop. Alternatively, the companion matrix
$\mathscr{C}_{\mathrm{nr}}$ given by Equation (\ref{eq:GyrCirMat})
and its characteristic polynomial as in Equation (\ref{eq:GyrCharPol2})
take the following forms
\begin{subequations}
\begin{equation}
\mathscr{C}_{\mathrm{nr}}=\left[\begin{array}{rrrr}
0 & 0 & 1 & 0\\
0 & 0 & 0 & 1\\
-\xi_{1} & 0 & 0 & \frac{R_{\mathrm{g}}}{L_{1}}\\
0 & -\xi_{2} & -\frac{R_{\mathrm{g}}}{L_{2}} & 0
\end{array}\right],\label{eq:GyrComp}
\end{equation}
\begin{equation}
\chi_{\mathrm{nr}}\left(h\right)=h^{2}+h\left(\xi_{1}+\xi_{2}+\frac{R_{\mathrm{g}}^{2}}{L_{1}L_{2}}\right)+\xi_{1}\xi_{2},\quad h=s^{2}.\label{eq:GyrCharPol4}
\end{equation}
\end{subequations}

Considering the degenerate eigenvalues satisfying equation $\chi_{\mathrm{nr}}\left(h\right)=0$,
we turn our attention to the discriminant of the quadratic polynomial
of Equation (\ref{eq:GyrCharPol4}),
\begin{equation}
\Delta_{\mathrm{nr},h}=\frac{R_{\mathrm{g}}^{4}}{L_{1}^{2}L_{2}^{2}}+\frac{2\left(\xi_{1}+\xi_{2}\right)R_{\mathrm{g}}^{2}}{L_{1}L_{2}}+\left(\xi_{1}-\xi_{2}\right)^{2},\label{eq:GyrDegCond1}
\end{equation}
where the solutions of the quadratic equation of Equation (\ref{eq:GyrCharPol4})
are
\begin{equation}
h_{\pm}=\frac{-\left(\xi_{1}+\xi_{2}+\frac{R_{g}^{2}}{L_{1}L_{2}}\right)\pm\sqrt{\Delta_{\mathrm{nr},h}}}{2}.\label{eq:GyrDegCond2}
\end{equation}
The corresponding four solutions $s$ of the characteristic polynomial
are given by
\begin{equation}
s=\pm\sqrt{h_{+}},\pm\sqrt{h_{-}},\label{eq:GyrDegCond3}
\end{equation}
In this case, the eigenvalue degeneracy condition becomes $\Delta_{\mathrm{nr},h}=0$.
It is possible to view degeneracy condition equation as a constraint
on the coefficients of the quadratic in polynomial $\chi_{\mathrm{nr}}\left(h\right)=0$
and consequently on the circuit parameters, namely
\begin{equation}
L_{1}^{2}L_{2}^{2}\Delta_{\mathrm{nr},h}=g^{2}+2\left(\xi_{1}+\xi_{2}\right)gL_{1}L_{2}+\left(\xi_{1}-\xi_{2}\right)^{2}L_{1}^{2}L_{2}^{2}=0,\quad g=R_{\mathrm{g}}^{2},\label{eq:GyrDegCond4}
\end{equation}
and we refer to positive $g$ as the \emph{gyration resistance square}.
Since $R_{\mathrm{g}}$ is real, then $g=R_{\mathrm{g}}^{2}$ is real
as well. Equation (\ref{eq:GyrDegCond4}) is a quadratic equation
for $g$, which has exactly two solutions of
\begin{equation}
g_{\mathrm{d,\pm}}=-L_{1}L_{2}\left(\sqrt{\xi_{1}}\mp\sqrt{\xi_{2}}\right)^{2},\label{eq:GyrDegCond6}
\end{equation}
which lead to two degenerate cases. For the two \textit{special value
of gyration resistance square} $g_{\mathrm{d,\pm}}$, we get two degenerate
roots for $h_{\pm}$ as

\begin{equation}
h_{\mathrm{d,}\pm}=-\frac{\xi_{1}+\xi_{2}+\frac{g_{\mathrm{d,\pm}}}{L_{1}L_{2}}}{2}=\mp\sqrt{\xi_{1}\xi_{2}}.
\end{equation}
Expression of $g_{\mathrm{d,\pm}}$ shown in Equation (\ref{eq:GyrDegCond6})
is real-valued if and only if $\xi_{1}\xi_{2}>0$, or equivalently
$\xi_{1}/\left|\xi_{1}\right|=\xi_{2}/\left|\xi_{2}\right|=\pm1$.
Also, the Jordan canonical form out of degeneracy condition, i.e.,
$\Delta_{\mathrm{nr},h}\neq0$, is expressed as

\begin{equation}
\mathscr{J}_{\mathrm{r}}=\left[\begin{array}{rrrr}
\mathrm{i}s_{1} & 0 & 0 & 0\\
0 & -\mathrm{i}s_{1} & 0 & 0\\
0 & 0 & \mathrm{i}s_{2} & 0\\
0 & 0 & 0 & -\mathrm{i}s_{2}
\end{array}\right],\quad\begin{array}{c}
s_{1}=\sqrt{\frac{\xi_{\mathrm{s}}+\sqrt{L_{1}^{2}L_{2}^{2}\Delta_{\mathrm{nr},h}}}{2L_{1}L_{2}}}\\
s_{2}=\sqrt{\frac{\xi_{\mathrm{s}}-\sqrt{L_{1}^{2}L_{2}^{2}\Delta_{\mathrm{nr},h}}}{2L_{1}L_{2}}}
\end{array},\quad\xi_{\mathrm{s}}=\left(\xi_{1}+\xi_{2}\right)L_{1}L_{2}+R_{\mathrm{g}}^{2},
\end{equation}
where $\pm\mathrm{i}s_{i}$ ($i=1,2$) are the non-degenerate eigenvalues
of the corresponding companion matrix. We write down eigenvectors
corresponding to the calculated non-degenerate eigenvalues as (when
$\Delta_{\mathrm{nr},h}\neq0$)

\begin{subequations}
\begin{equation}
\mathscr{V}_{\mathrm{nr}}=\left[\mathsf{e}_{\mathrm{nr},+s_{1}}|\mathsf{e}_{\mathrm{nr},-s_{1}}|\mathsf{e}_{\mathrm{nr},+s_{2}}|\mathsf{e}_{\mathrm{nr},-s_{2}}\right],
\end{equation}

\begin{equation}
\begin{array}{c}
\mathsf{e}_{\mathrm{nr},\pm s_{i}}=\left[\begin{array}{r}
\frac{2R_{\mathrm{g}}^{2}+2L_{1}L_{2}\xi_{2}-\xi_{\mathrm{s}}-\sqrt{L_{1}^{2}L_{2}^{2}\Delta_{\mathrm{nr},h}}}{2L_{1}R_{\mathrm{g}}\xi_{1}}\\
\mp\frac{L_{1}L_{2}\left(-s_{i}^{2}\right)^{3/2}+\xi_{\mathrm{s}}\sqrt{-s_{i}^{2}}}{L_{1}L_{2}\xi_{1}\xi_{2}}\\
\pm\frac{L_{1}L_{2}\left(-s_{i}^{2}\right)^{3/2}+\left(R_{\mathrm{g}}^{2}+L_{1}L_{2}\xi_{2}\right)\sqrt{-s_{i}^{2}}}{L_{1}R_{\mathrm{g}}\xi_{1}}\\
1
\end{array}\right].\end{array}
\end{equation}
\end{subequations}

There were two values for the degenerate gyration resistance square
as shown in Equation (\ref{eq:GyrDegCond6}), resulting in degenerate
purely imaginary or purely real eigenvalues. Firstly, if we consider
$g=g_{\mathrm{d},+}=-L_{1}L_{2}\left(\sqrt{\xi_{1}}-\sqrt{\xi_{2}}\right)^{2}$,
the degenerate characteristic polynomial is given by

\begin{equation}
\chi_{\mathrm{nr,+}}\left(s\right)=s^{4}+2s^{2}\sqrt{\xi_{1}\xi_{2}}+\xi_{1}\xi_{2}=\left(s^{2}+\sqrt{\xi_{1}\xi_{2}}\right)^{2}=0.
\end{equation}
Then, the corresponding degenerate companion matrix $\mathscr{C}_{\mathrm{nr,+}}$
by substituting $g=g_{\mathrm{d},+}$ is rewritten as

\begin{equation}
\mathscr{C}_{\mathrm{nr,+}}=\left[\begin{array}{rrrr}
0 & 0 & 1 & 0\\
0 & 0 & 0 & 1\\
-\xi_{1} & 0 & 0 & \sqrt{\frac{g_{\mathrm{d},+}}{L_{1}^{2}}}\\
0 & -\xi_{2} & -\sqrt{\frac{g_{\mathrm{d},+}}{L_{2}^{2}}} & 0
\end{array}\right].
\end{equation}
\[
.
\]
The Jordan canonical form of the companion matrix $\mathscr{C}_{\mathrm{nr,+}}$
with purely imaginary eigenvalues (purely real eigenfrequencies) is
described by

\begin{equation}
\mathscr{J}_{\mathrm{nr,+}}=\left[\begin{array}{rrrr}
\mathrm{i}s_{\mathrm{e}} & 1 & 0 & 0\\
0 & \mathrm{i}s_{\mathrm{e}} & 0 & 0\\
0 & 0 & -\mathrm{i}s_{\mathrm{e}} & 1\\
0 & 0 & 0 & -\mathrm{i}s_{\mathrm{e}}
\end{array}\right],\quad s_{\mathrm{e}}=\left(\xi_{\mathrm{e}}\right)^{\frac{1}{2}}=\left(\xi_{1}\xi_{2}\right)^{\frac{1}{4}}.
\end{equation}
Accordingly, the Jordan basis of the companion matrix $\mathscr{C}_{\mathrm{nr,+}}$
is calculated as

\begin{equation}
\mathscr{Z}_{\mathrm{nr,+}}=\left[\begin{array}{rrrr}
\frac{\sqrt{\xi_{1}\xi_{2}}-\xi_{1}}{4\mathrm{i}\sqrt{\xi_{\mathrm{e}}}} & \frac{1}{2} & -\frac{\sqrt{\xi_{1}\xi_{2}}-\xi_{1}}{4\mathrm{i}\sqrt{\xi_{\mathrm{e}}}} & \frac{1}{2}\\
-\frac{\xi_{1}\sqrt{L_{1}L_{2}\left(2\xi_{\mathrm{e}}-\xi_{1}-\xi_{2}\right)}}{4L_{2}\xi_{\mathrm{e}}} & -\mathrm{i}\frac{\xi_{1}\sqrt{L_{1}L_{2}\left(2\xi_{\mathrm{e}}-\xi_{1}-\xi_{2}\right)}}{4L_{2}\xi_{\mathrm{e}}\sqrt{\xi_{\mathrm{e}}}} & -\frac{\xi_{1}\sqrt{L_{1}L_{2}\left(2\xi_{\mathrm{e}}-\xi_{1}-\xi_{2}\right)}}{4L_{2}\xi_{\mathrm{e}}} & \mathrm{i}\frac{\xi_{1}\sqrt{L_{1}L_{2}\left(2\xi_{\mathrm{e}}-\xi_{1}-\xi_{2}\right)}}{4L_{2}\xi_{\mathrm{e}}\sqrt{\xi_{\mathrm{e}}}}\\
\frac{\xi_{1}\left(\xi_{2}-\xi_{\mathrm{e}}\right)}{4\xi_{\mathrm{e}}} & -\frac{\xi_{1}\left(\xi_{2}+\xi_{\mathrm{e}}\right)}{4\mathrm{i}\xi_{\mathrm{e}}\sqrt{\xi_{\mathrm{e}}}} & \frac{\xi_{1}\left(\xi_{2}-\xi_{\mathrm{e}}\right)}{4\xi_{\mathrm{e}}} & \frac{\xi_{1}\left(\xi_{2}+\xi_{\mathrm{e}}\right)}{4\mathrm{i}\xi_{\mathrm{e}}\sqrt{\xi_{\mathrm{e}}}}\\
\frac{\xi_{1}\sqrt{L_{1}L_{2}\left(2\xi_{\mathrm{e}}-\xi_{1}-\xi_{2}\right)}}{4\mathrm{i}L_{2}\sqrt{\xi_{\mathrm{e}}}} & 0 & -\frac{\xi_{1}\sqrt{L_{1}L_{2}\left(2\xi_{\mathrm{e}}-\xi_{1}-\xi_{2}\right)}}{4\mathrm{i}L_{2}\sqrt{\xi_{\mathrm{e}}}} & 0
\end{array}\right].
\end{equation}

Secondly, if we consider $g=g_{\mathrm{d},-}=-L_{1}L_{2}\left(\xi_{1}+\xi_{2}+2\sqrt{\xi_{1}\xi_{2}}\right)$,
the degenerate characteristic polynomial is rewritten as

\begin{equation}
\chi_{\mathrm{nr,-}}\left(s\right)=s^{4}-2s^{2}\sqrt{\xi_{1}\xi_{2}}+\xi_{1}\xi_{2}=\left(s^{2}-\sqrt{\xi_{1}\xi_{2}}\right)^{2}=0.
\end{equation}
Then, the degenerate corresponding companion matrix $\mathscr{C}_{\mathrm{nr,-}}$
is rewritten as

\begin{equation}
\mathscr{C}_{\mathrm{nr,-}}=\left[\begin{array}{rrrr}
0 & 0 & 1 & 0\\
0 & 0 & 0 & 1\\
-\xi_{1} & 0 & 0 & -\sqrt{\frac{g_{\mathrm{d},-}}{L_{1}^{2}}}\\
0 & -\xi_{2} & \sqrt{\frac{g_{\mathrm{d},-}}{L_{2}^{2}}} & 0
\end{array}\right].
\end{equation}
The Jordan canonical form of the companion matrix $\mathscr{C}_{\mathrm{nr,-}}$
with purely real eigenvalues (purely imaginary eigenfrequencies) is
expressed by

\begin{equation}
\mathscr{J}_{\mathrm{nr,-}}=\left[\begin{array}{rrrr}
s_{\mathrm{e}} & 1 & 0 & 0\\
0 & s_{\mathrm{e}} & 0 & 0\\
0 & 0 & -s_{\mathrm{e}} & 1\\
0 & 0 & 0 & -s_{\mathrm{e}}
\end{array}\right].
\end{equation}
Finally, the Jordan basis of the companion matrix $\mathscr{C}_{\mathrm{nr,-}}$
is obtained as

\begin{equation}
\mathscr{Z}_{\mathrm{nr,-}}=\left[\begin{array}{rrrc}
-\frac{\sqrt{\xi_{1}\xi_{2}}+\xi_{1}}{4\sqrt{\xi_{\mathrm{e}}}} & \frac{1}{2} & \frac{\sqrt{\xi_{1}\xi_{2}}+\xi_{1}}{4\sqrt{\xi_{\mathrm{e}}}} & \frac{1}{2}\\
\mathrm{i}\frac{\xi_{1}\sqrt{L_{1}L_{2}\left(2\xi_{\mathrm{e}}+\xi_{1}+\xi_{2}\right)}}{4L_{2}\xi_{\mathrm{e}}} & -\mathrm{i}\frac{\xi_{1}\sqrt{L_{1}L_{2}\left(2\xi_{\mathrm{e}}+\xi_{1}+\xi_{2}\right)}}{4L_{2}\xi_{\mathrm{e}}\sqrt{\xi_{\mathrm{e}}}} & \mathrm{i}\frac{\xi_{1}\sqrt{L_{1}L_{2}\left(2\xi_{\mathrm{e}}+\xi_{1}+\xi_{2}\right)}}{4L_{2}\xi_{\mathrm{e}}} & \mathrm{i}\frac{\xi_{1}\sqrt{L_{1}L_{2}\left(2\xi_{\mathrm{e}}+\xi_{1}+\xi_{2}\right)}}{4L_{2}\xi_{\mathrm{e}}\sqrt{\xi_{\mathrm{e}}}}\\
-\frac{\xi_{1}\left(\xi_{2}+\xi_{\mathrm{e}}\right)}{4\xi_{\mathrm{e}}} & \frac{\xi_{1}\left(\xi_{2}-\xi_{\mathrm{e}}\right)}{4\xi_{\mathrm{e}}\sqrt{\xi_{\mathrm{e}}}} & -\frac{\xi_{1}\left(\xi_{2}+\xi_{\mathrm{e}}\right)}{4\xi_{\mathrm{e}}} & -\frac{\xi_{1}\left(\xi_{2}-\xi_{\mathrm{e}}\right)}{4\xi_{\mathrm{e}}\sqrt{\xi_{\mathrm{e}}}}\\
\mathrm{i}\frac{\xi_{1}\sqrt{L_{1}L_{2}\left(2\xi_{\mathrm{e}}+\xi_{1}+\xi_{2}\right)}}{4L_{2}\sqrt{\xi_{\mathrm{e}}}} & 0 & -\mathrm{i}\frac{\xi_{1}\sqrt{L_{1}L_{2}\left(2\xi_{\mathrm{e}}+\xi_{1}+\xi_{2}\right)}}{4L_{2}\sqrt{\xi_{\mathrm{e}}}} & 0
\end{array}\right].
\end{equation}

The fundamental of Lagrangian for a linear system is explained in
Subsection \ref{subsec:RecipLag} and the associated coefficients
for the GNC are given by

\begin{equation}
\alpha_{\mathrm{nr}}=\left[\begin{array}{rr}
L_{1} & 0\\
0 & L_{2}
\end{array}\right],\quad\theta_{\mathrm{nr}}=\left[\begin{array}{rr}
0 & -\frac{R_{\mathrm{g}}}{2}\\
\frac{R_{\mathrm{g}}}{2} & 0
\end{array}\right],\quad\eta_{\mathrm{nr}}=\left[\begin{array}{rr}
\frac{1}{C_{1}} & 0\\
0 & \frac{1}{C_{2}}
\end{array}\right].
\end{equation}

It follows from Equation (\ref{eq:Lag4}) that the necessary and sufficient
condition of the nonreciprocity is $\theta_{\mathrm{nr}}\neq\theta_{\mathrm{nr}}^{\mathrm{T}}$.
Then, the Lagrangian equation of GNC is written as follows:

\begin{subequations}
\begin{equation}
\mathcal{L}_{\mathrm{nr}}\left(\mathrm{q}_{\mathrm{nr}},\partial_{t}\mathrm{q}_{\mathrm{nr}}\right)=\frac{1}{2}\left[\begin{array}{r}
\mathrm{q}_{\mathrm{nr}}\\
\partial_{t}\mathrm{q}_{\mathrm{nr}}
\end{array}\right]^{\mathrm{T}}\mathrm{M}_{\mathrm{L,nr}}\left[\begin{array}{r}
\mathrm{q}_{\mathrm{nr}}\\
\partial_{t}\mathrm{q}_{\mathrm{nr}}
\end{array}\right],
\end{equation}
\begin{equation}
\mathrm{M}_{\mathrm{L,nr}}=\left[\begin{array}{rrrr}
-\frac{1}{C_{1}} & 0 & 0 & \frac{\mathit{R_{\mathrm{g}}}}{2}\\
0 & -\frac{1}{C_{2}} & -\frac{\mathit{R_{\mathrm{g}}}}{2} & 0\\
0 & -\frac{\mathit{R_{\mathrm{g}}}}{2} & L_{1} & 0\\
\frac{\mathit{R_{\mathrm{g}}}}{2} & 0 & 0 & L_{2}
\end{array}\right].
\end{equation}
\end{subequations}
Moreover, the Hamiltonian formulation can also be used to show the
GNC characteristics (see Subsection \ref{subsec:Ham} for more information).
Then, the $4\times4$ matrix $\mathrm{M}_{\mathrm{H,nr}}$ required
for the Hamiltonian formulation is given by
\begin{gather}
\mathrm{M}_{\mathrm{H,nr}}=\left[\begin{array}{rrrr}
\frac{R_{\mathrm{g}}^{2}}{4L_{2}}+\frac{1}{C_{1}} & 0 & 0 & -\frac{R_{\mathrm{g}}}{2L_{2}}\\
0 & \frac{R_{\mathrm{g}}^{2}}{4L_{1}}+\frac{1}{C_{2}} & \frac{R_{\mathrm{g}}}{2L_{1}} & 0\\
0 & \frac{\mathit{R_{\mathrm{g}}}}{2L_{1}} & \frac{1}{L_{1}} & 0\\
-\frac{R_{\mathrm{g}}}{2L_{2}} & 0 & 0 & \frac{1}{L_{2}}
\end{array}\right].\label{eq:dlag7a-1}
\end{gather}
Ultimately, the Hamiltonian matrix for the GNC is given by by
\begin{equation}
\mathscr{H}_{\mathrm{nr}}=\mathbb{J}\mathrm{M}_{\mathrm{H,nr}}=\left[\begin{array}{rrrr}
0 & \frac{\mathit{R_{\mathrm{g}}}}{2L_{1}} & \frac{1}{L_{1}} & 0\\
-\frac{\mathit{R_{\mathrm{g}}}}{2L_{2}} & 0 & 0 & \frac{1}{L_{2}}\\
-\frac{\mathit{R_{\mathrm{g}}^{2}}}{4L_{2}}-\frac{1}{C_{1}} & 0 & 0 & \frac{\mathit{R_{\mathrm{g}}}}{2L_{2}}\\
0 & -\frac{\mathit{R_{\mathrm{g}}^{2}}}{4L_{1}}-\frac{1}{C_{2}} & -\frac{\mathit{R_{\mathrm{g}}}}{2L_{1}} & 0
\end{array}\right].\label{eq:MJHam1a}
\end{equation}

According to Equation (\ref{eq:Hamil1b}), we can see that in the
nonreciprocal circuit with $\theta_{\mathrm{nr}}\neq0$, the relation
between the \textit{momentum} $\mathrm{p}_{\mathrm{nr}}$ and the
\textit{current} $\partial_{t}\mathrm{q}_{\mathrm{nr}}$ also depends
on the charge $\mathrm{q}_{\mathrm{nr}}$. In addition, it is important
to point out that a circuit with gyrators does not necessarily guarantee
nonreciprocity, but it \textit{can} lead to nonreciprocity.

\section{Analysis of Reciprocity and Nonreciprocity\label{sec:Analysis-of-Reciprocity}}

In this section, we analyze reciprocity properties in both reciprocal
and nonreciprocal circuits. This particular feature will be explored
in the Lagrangian and eigenvectors of both PRC and GNC.

\subsection{Lagrangian\label{subsec:ReciprocityLageq}}

According to investigation provided in Section \ref{sec:Principal-Reciprocal-Circuit},
the Lagrangian associated with the PRC is depicted as

\begin{equation}
\left.\mathcal{L}_{\mathrm{r}}\left(\mathrm{q}_{\mathrm{r}},\partial_{t}\mathrm{q}_{\mathrm{r}}\right)\right|_{\Theta_{\mathrm{r}}}=\frac{L_{1}\left(\partial_{t}q_{1}\right)^{2}}{2}+\frac{L_{0}\left(\partial_{t}q_{0}\right)^{2}}{2}-\frac{\left(q_{1}\right)^{2}}{2C_{1}}-\frac{\left(q_{1}-q_{0}\right)^{2}}{2C_{0}}.\label{eq: RecipLag2}
\end{equation}
An arrow of time is a concept that proposes the \textquotedbl one-way
direction\textquotedbl{} or \textquotedbl asymmetry\textquotedbl{}
of time. If we change the direction of arrow of time, the Lagrangian
associated with PRC is expressed by

\begin{equation}
\left.\mathcal{L}_{\mathrm{r}}\left(\mathrm{q}_{\mathrm{r}},-\partial_{t}\mathrm{q}_{\mathrm{r}}\right)\right|_{\Theta_{\mathrm{r}}}=\frac{L_{1}\left(\partial_{t}q_{1}\right)^{2}}{2}+\frac{L_{0}\left(\partial_{t}q_{0}\right)^{2}}{2}-\frac{\left(q_{1}\right)^{2}}{2C_{1}}-\frac{\left(q_{1}-q_{0}\right)^{2}}{2C_{0}}.\label{eq: RecipLagNeg}
\end{equation}
By comparing Equations (\ref{eq: RecipLag2}) and (\ref{eq: RecipLagNeg})
we have

\begin{equation}
\left.\mathcal{L}_{\mathrm{r}}\left(\mathrm{q}_{\mathrm{r}},\partial_{t}\mathrm{q}_{\mathrm{r}}\right)\right|_{\Theta_{\mathrm{r}}}=\left.\mathcal{L}_{\mathrm{r}}\left(\mathrm{q}_{\mathrm{r}},-\partial_{t}\mathrm{q}_{\mathrm{r}}\right)\right|_{\Theta_{\mathrm{r}}}.
\end{equation}
In the above equation, we can see that there is symmetry in time,
which refers to reciprocity within the PRC (see Definition \ref{def: LagrangianRec}).
\begin{defn}[Lagrangian of reciprocal and nonreciprocal systems\label{def: LagrangianRec}]
For a system described by coordinates $F=\left.\left[f_{i}\right]\right|_{i=1}^{n}$
and time $t$, the time reversal symmetry can be formulated as follows.
For any trajectory $F\left(t\right)$ of the system, $\tilde{F}\left(t\right)=F\left(-t\right)$
is also its trajectory. In terms of the Lagrangian function $\mathcal{L}=\mathcal{L}\left(F,\dot{F}\right)$,
it means the invariance of Lagrangian function under the transformation
$\dot{F}\rightarrow-\dot{F}$ \cite{figotin2001spectra}:

\begin{equation}
\mathcal{L}\left(F,\dot{F}\right)=\mathcal{L}\left(F,-\dot{F}\right).
\end{equation}
Also, for a system with broken time reversal symmetry we observe $\mathcal{L}\left(F,\dot{F}\right)\neq\mathcal{L}\left(F,-\dot{F}\right)$.
\end{defn}

On the other hand, based on the expression given in Section \ref{sec:Gyrator-based-Circuit},
the Lagrangian associated with GNC is expressed by

\begin{equation}
\left.\mathcal{L}_{\mathrm{nr}}\left(\mathrm{q}_{\mathrm{nr}},\partial_{t}\mathrm{q}_{\mathrm{nr}}\right)\right|_{\Theta_{\mathrm{nr}}}=\frac{L_{1}\left(\partial_{t}q_{1}\right)^{2}}{2}+\frac{L_{2}\left(\partial_{t}q_{2}\right)^{2}}{2}-\frac{\left(q_{1}\right)^{2}}{2C_{1}}-\frac{\left(q_{2}\right)^{2}}{2C_{2}}+\frac{R_{\mathrm{g}}\left(q_{1}\partial_{t}q_{2}-q_{2}\partial_{t}q_{1}\right)}{2}.\label{eq:GyrLag2}
\end{equation}
Then, by inverting the direction of the time we have

\begin{equation}
\left.\mathcal{L}_{\mathrm{nr}}\left(\mathrm{q}_{\mathrm{nr}},-\partial_{t}\mathrm{q}_{\mathrm{nr}}\right)\right|_{\Theta_{\mathrm{nr}}}=\frac{L_{1}\left(\partial_{t}q_{1}\right)^{2}}{2}+\frac{L_{2}\left(\partial_{t}q_{2}\right)^{2}}{2}-\frac{\left(q_{1}\right)^{2}}{2C_{1}}-\frac{\left(q_{2}\right)^{2}}{2C_{2}}-\frac{R_{\mathrm{g}}\left(q_{1}\partial_{t}q_{2}-q_{2}\partial_{t}q_{1}\right)}{2},\label{eq:GyrLagNeg}
\end{equation}
where we observe the sign of nonreciprocity in the last term of the
above equation. Now, by inverting the direction of gyration resistance
($R_{\mathrm{g}}\rightarrow-R_{\mathrm{g}}$) in addition to reversing
the direction of time, we rewrite the Lagrangian associated with the
new circuit by using a new set of parameters $\widetilde{\Theta}_{\mathrm{nr}}=\left\{ L_{1},L_{2},C_{1},C_{2},-R_{\mathrm{g}}\right\} $
as

\begin{equation}
\left.\mathcal{L}_{\mathrm{nr}}\left(\mathrm{q}_{\mathrm{nr}},-\partial_{t}\mathrm{q}_{\mathrm{nr}}\right)\right|_{\widetilde{\Theta}_{\mathrm{nr}}}=\frac{L_{1}\left(\partial_{t}q_{1}\right)^{2}}{2}+\frac{L_{2}\left(\partial_{t}q_{2}\right)^{2}}{2}-\frac{\left(q_{1}\right)^{2}}{2C_{1}}-\frac{\left(q_{2}\right)^{2}}{2C_{2}}+\frac{R_{\mathrm{g}}\left(q_{1}\partial_{t}q_{2}-q_{2}\partial_{t}q_{1}\right)}{2}.\label{eq:GyrLagNegRgNeg}
\end{equation}
Finally, by comparing Equations (\ref{eq:GyrLag2}), (\ref{eq:GyrLagNeg})
and (\ref{eq:GyrLagNegRgNeg}), we have

\begin{equation}
\left.\mathcal{L}_{\mathrm{nr}}\left(\mathrm{q}_{\mathrm{nr}},\partial_{t}\mathrm{q}_{\mathrm{nr}}\right)\right|_{\Theta_{\mathrm{nr}}}=\left.\mathcal{L}_{\mathrm{nr}}\left(\mathrm{q}_{\mathrm{nr}},-\partial_{t}\mathrm{q}_{\mathrm{nr}}\right)\right|_{\widetilde{\Theta}_{\mathrm{nr}}}\neq\left.\mathcal{L}_{\mathrm{nr}}\left(\mathrm{q}_{\mathrm{nr}},-\partial_{t}\mathrm{q}_{\mathrm{nr}}\right)\right|_{\Theta_{\mathrm{nr}}},
\end{equation}
which demonstrate the nonreciprocity in the GNC (see Definition \ref{def: LagrangianRec}).
In spite of this, the circuit with the same Lagrangian is achieved
by simultaneously inverting the direction of the time and gyration
resistance. Alternately, by removing the source of nonreciprocity
in the circuit, i.e., the gyrator, the circuit becomes a reciprocal
circuit composed of two uncoupled LC resonators. The properties of
a single LC resonator are discussed in Appendix \ref{sec:Single-LC-Circuit}.

\subsection{Eigenvectors}

As explained in Subsection \ref{subsec:Recip-Eigenvectors}, the eigenvectors
associated to the PRC in the general form are expressed as

\begin{equation}
\begin{array}{c}
\mathsf{e}_{\mathrm{r,}\pm s_{i}}=\left[\begin{array}{c}
\mathsf{q}_{\pm s_{i}}\\
\dot{\mathsf{q}}_{\pm s_{i}}
\end{array}\right]\stackrel{\mathrm{def}}{=}\left[\begin{array}{r}
\mp\frac{\left(-s_{i}^{2}\right)^{3/2}+\left(\xi_{0}+\delta\right)\sqrt{-s_{i}^{2}}}{\xi_{0}\xi_{1}}\\
\mp\frac{\left(-s_{i}^{2}\right)^{3/2}+\xi_{\mathrm{s}}\sqrt{-s_{i}^{2}}}{\xi_{0}\xi_{i}}\\
1-\frac{s_{i}^{2}}{\xi_{0}}\\
1
\end{array}\right],\quad\xi_{\mathrm{s}}=\xi_{1}+\xi_{0}+\delta,\end{array}\quad i=1,2.\label{eq:RecipEigVec}
\end{equation}
Due to reciprocity, we observe the change in the sign of charges when
we inverse the sign of $s_{i}$ ($i=1,2$), i.e., $\mathsf{q}_{+s_{i}}=-\mathsf{q}_{-s_{i}}$
($i=1,2$), whereas the sign of the first derivative of charges remains
constant, i.e., $\dot{\mathsf{q}}_{+s_{i}}=\dot{\mathsf{q}}_{-s_{i}}$
($i=1,2$). The reciprocity feature of PRC is evident from these properties.

Next, we study the GNC eigenvectors to explore a reciprocity feature.
Based on the presented expression in Section \ref{sec:Gyrator-based-Circuit},
the eigenvectors associated to the GNC are presented as

\begin{equation}
\begin{array}{c}
\mathsf{e}_{\mathrm{nr,}\pm s_{i}}=\left[\begin{array}{c}
\mathsf{q}_{\pm s_{i}}\\
\dot{\mathsf{q}}_{\pm s_{i}}
\end{array}\right]\stackrel{\mathrm{def}}{=}\left[\begin{array}{r}
\frac{2R_{\mathrm{g}}^{2}+2L_{1}L_{2}\xi_{2}-\xi_{\mathrm{s}}-\sqrt{L_{1}^{2}L_{2}^{2}\Delta_{\mathrm{nr},h}}}{2L_{1}R_{\mathrm{g}}\xi_{1}}\\
\mp\frac{L_{1}L_{2}\left(-s_{i}^{2}\right)^{3/2}+\xi_{\mathrm{s}}\sqrt{-s_{i}^{2}}}{L_{1}L_{2}\xi_{1}\xi_{2}}\\
\pm\frac{L_{1}L_{2}\left(-s_{i}^{2}\right)^{3/2}+\left(R_{\mathrm{g}}^{2}+L_{1}L_{2}\xi_{2}\right)\sqrt{-s_{i}^{2}}}{L_{1}R_{\mathrm{g}}\xi_{1}}\\
1
\end{array}\right],\quad\xi_{\mathrm{s}}=\left(\xi_{1}+\xi_{2}\right)L_{1}L_{2}+R_{\mathrm{g}}^{2},\quad i=1,2.\end{array}\label{eq:GyrEigVec}
\end{equation}
We cannot observe the same behavior for the sign of charges and their
first derivative from the above expression for the eigenvectors of
the GNC, i.e., $\mathsf{q}_{+s_{i}}\neq-\mathsf{q}_{-s_{i}}$, $\dot{\mathsf{q}}_{+s_{i}}\neq\dot{\mathsf{q}}_{-s_{i}}$($i=1,2$).
In light of these eigenvector properties, it can be concluded that
GNC is not reciprocal.
\begin{rem}[Nonreciprocity and symmetries of the set of eigenvectors]
Our studies indicate that nonreciprocity is manifested in the breakdown
of natural symmetries of the set of eigenvectors rather than eigenvalues.
Indeed, the Lagrangian and eigenvectors of a circuit are capable of
reflecting the reciprocity properties of the circuit. Despite this,
nonreciprocity cannot be captured in the Jordan canonical form and
consequently in the eigenvalues of the circuit matrix.

\end{rem}

We demonstrated that the Jordan canonical form of the circuit does
not reflect reciprocity. Consequently, despite the fact that reciprocal
and nonreciprocal circuits differ from many perspectives, these two
circuits with different topologies can produce the same Jordan canonical
form under certain conditions.

\section{Reciprocal and Nonreciprocal Circuits Transformation\label{sec:Equivalency}}

\subsection{Equivalency condition in characteristic polynomial}

By comparing the characteristic polynomial of the PRC expressed in
Equation (\ref{eq:RecipCharEq}) and the characteristic polynomial
of the GNC stated in Equation (\ref{eq:GyrCharPol3}) we observe the
conditions to obtain equivalency between the characteristic polynomials
of these two circuits by equating coefficients as

\begin{equation}
\left\{ \begin{array}{l}
\left\{ s^{1}\right\} :\frac{R_{\mathrm{g}}^{2}}{L_{1}L_{2}}+\frac{1}{L_{2}C_{2}}+\frac{1}{L_{1}C_{1}}=\frac{1}{L_{1}C_{0}}+\frac{1}{L_{0}C_{0}}+\frac{1}{L_{1}C_{1}}\\
\left\{ s^{0}\right\} :\frac{1}{L_{1}C_{1}L_{2}C_{2}}=\frac{1}{L_{0}C_{0}L_{1}C_{1}}
\end{array}\right.
\end{equation}
Then, we suppose that the first fundamental loop (that include $L_{1}$
and $C_{1}$) in both circuits has the same resonance frequency. So,
the conditions to have the same Jordan canonical form for both PRC
and GNC are summarized as follows:

\begin{equation}
R_{\mathrm{g}}^{2}=R_{\mathrm{g,eq}}^{2}=\frac{L_{2}}{C_{0}}=\frac{L_{0}}{C_{2}},\label{eq:EqCond1}
\end{equation}
where $R_{\mathrm{g,eq}}$ is the \textit{equivalency value of gyration
resistance} that can be used to get the same eigenfrequencies in PRC
and GNC.

\subsection{Equivalent circuit representation}

\begin{figure}
\begin{centering}
\includegraphics[width=0.5\textwidth]{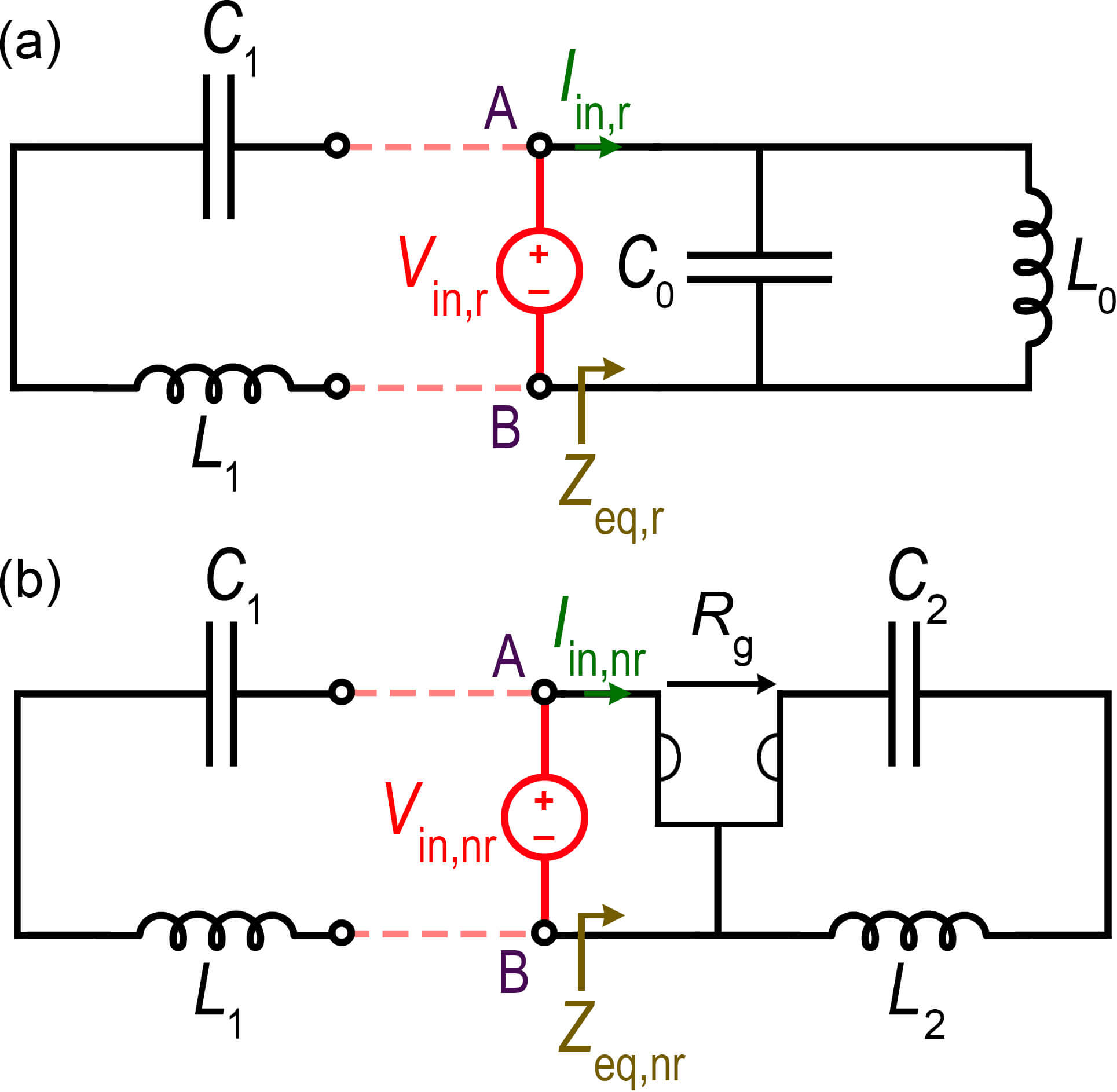}
\par\end{centering}
\caption{The equivalent circuit and the condition for obtaining (a) reciprocal
and (b) nonreciprocal circuits with the same Jordan canonical forms
and consequently the same eigenfrequencies by selecting the equivalency
value for the gyration resistance $R_{\mathrm{g,eq}}$. The right
side of both circuits can have the same equivalent impedance value
if $R_{\mathrm{g,eq}}=\sqrt{L_{0}/C_{2}}=\sqrt{L_{2}/C_{0}}$. The
left loop in both circuits includes a series resonator with the same
resonance frequency.\label{fig:Equivalent Circuit}}
\end{figure}
By using the equivalent circuit representation that can be observed
from two selected points of a desired network, we can determine the
equivalency condition between reciprocal and nonreciprocal circuits
\cite{wu2004theory,izmailian2013two}. As originally described in
network theory, Thévenin's theorem asserts that \textquotedblleft \textit{for
any linear electrical network with only voltage and current sources,
and impedances can be substituted at input ports by using a combination
of an equivalent impedance $Z_{\mathrm{Th}}$ in a series connection
with an equivalent voltage source $V_{\mathrm{Th}}$}''. The equivalent
impedance $Z_{\mathrm{Th}}$ is the impedance observed from the input
port if all ideal current sources in the circuit were substituted
by an open circuit and all ideal voltage sources were substituted
by a short circuit. Also, the equivalent voltage $V_{\mathrm{Th}}$
is the voltage calculated at the input port of the circuit when the
input port is open. In the case of a short circuit at the input port,
the current flowing from the short circuit would be \textit{$I_{\mathrm{Th}}=V_{\mathrm{Th}}/Z_{\mathrm{Th}}$}.
As a result, equivalent impedance \textit{$Z_{\mathrm{Th}}$} could
also be calculated as $Z_{\mathrm{Th}}=V_{\mathrm{Th}}/I_{\mathrm{Th}}$.
To simplify circuit analysis, Thévenin's theorem and its dual, Norton's
theorem, are often used. By using Thévenin's theorem, any circuit's
sources and impedances can be converted to a Thévenin equivalents;
in some cases, this may be more convenient than Kirchhoff's voltage
and current laws.

According to Figure \ref{fig:Equivalent Circuit}(a), we cut the PRC
circuit into two segments and characterize the right segment using
the Thévenin's equivalent circuit. The input impedance $Z_{\mathrm{eq,r}}$,
that is identical to the equivalent Thévenin's impedance\textit{ }observed
from port AB, for the PRC is calculated by

\begin{equation}
Z_{\mathrm{eq,r}}=\frac{V_{\mathrm{in,r}}}{I_{\mathrm{in,r}}}=\frac{1}{j\omega C_{0}+\frac{1}{j\omega L_{0}}}.
\end{equation}
By using the same approach for the GNC shown in Figure \ref{fig:Equivalent Circuit}(b),
we calculate equivalent input impedance $Z_{\mathrm{eq,nr}}$ as

\begin{equation}
Z_{\mathrm{eq,nr}}=\frac{V_{\mathrm{in,nr}}}{I_{\mathrm{in,nr}}}=\frac{R_{\mathrm{g}}^{2}}{j\omega L_{2}+\frac{1}{j\omega C_{2}}}.
\end{equation}
Then, if we consider the equality of input impedances $Z_{\mathrm{eq,r}}=Z_{\mathrm{eq,nr}}$,
we write

\begin{equation}
R_{\mathrm{g}}^{2}=R_{\mathrm{g,eq}}^{2}=\frac{\omega^{2}L_{2}C_{2}L_{0}-L_{0}}{\omega^{2}C_{0}C_{2}L_{0}-C_{2}}=\frac{L_{0}\left(\omega^{2}L_{2}C_{2}-1\right)}{C_{2}\left(\omega^{2}L_{0}C_{0}-1\right)}.\label{eq:EquivCond1}
\end{equation}
In order to have a frequency-independent value for gyration resistance
$R_{\mathrm{g}}$, we apply the following condition:

\begin{equation}
L_{2}C_{2}=L_{0}C_{0},\:\text{or equivalently}\:\xi_{2}=\xi_{0}.\label{eq:EquivCond2}
\end{equation}
Consequently, from Equations (\ref{eq:EquivCond1}) and (\ref{eq:EquivCond1}),
the \textit{equivalency value of gyration resistance} is given by

\begin{equation}
R_{\mathrm{g,eq}}^{2}=\frac{L_{2}}{C_{0}}=\frac{L_{0}}{C_{2}}.\label{eq:EqCond2}
\end{equation}
The condition shown in Equation (\ref{eq:EqCond1}) is consistent
with the condition derived in Equation (\ref{eq:EqCond2}) by applying
a different approach.

\section{Example with Circuit Simulator Numerical Results\label{sec:Circuit-Simulator-Simulation}}

\begin{figure}
\begin{centering}
\includegraphics[width=0.82\textwidth]{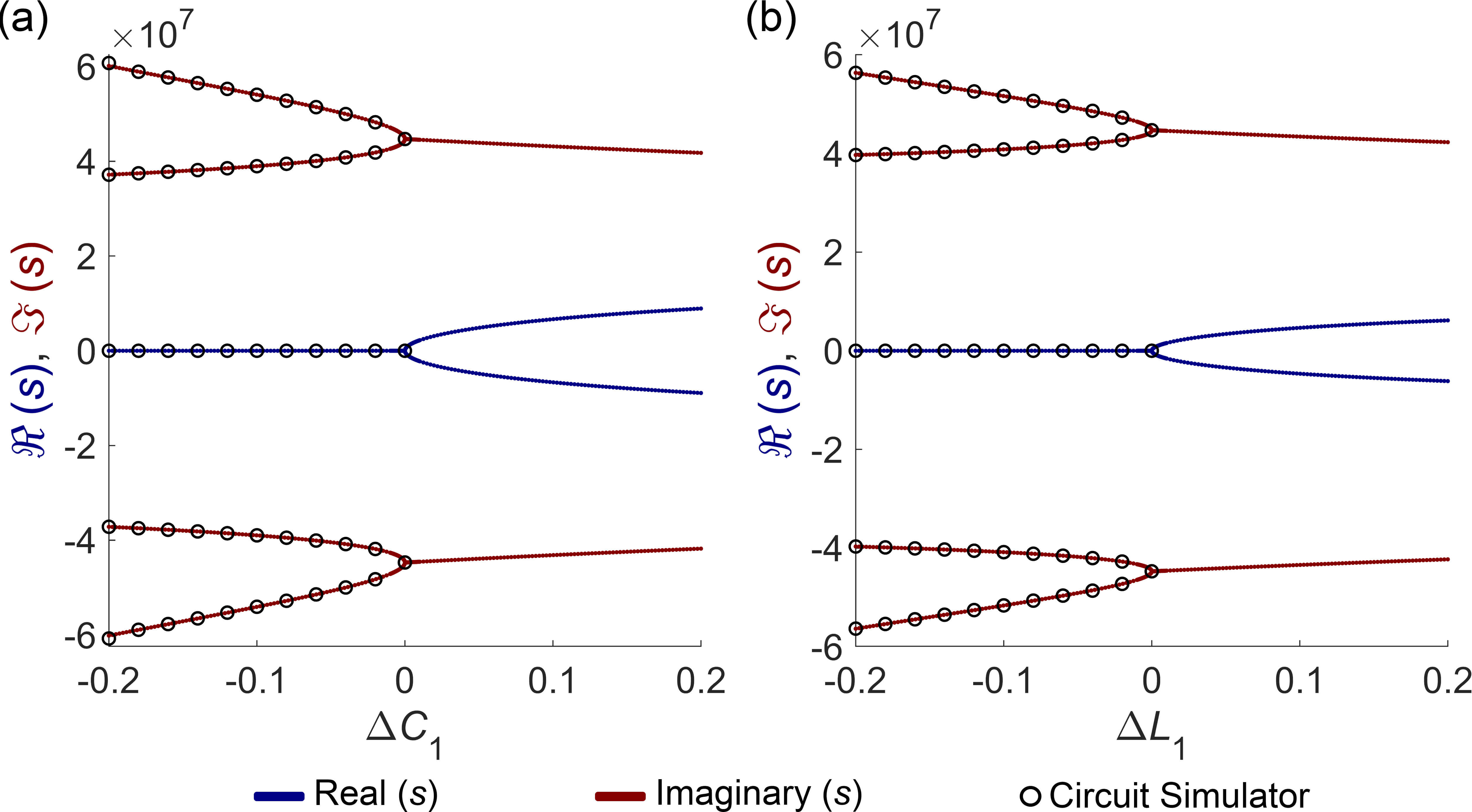}
\par\end{centering}
\caption{The real (dark blue solid curve) and imaginary (dark red solid curve)
parts of the four eigenvalues as in Equation (\ref{eq:RecipCharEq})
assuming that (a) $\Delta L_{0}=\Delta L_{1}=\Delta C_{0}=0$, and
$\Delta C_{1}$ varies and (b) $\Delta L_{0}=\Delta C_{0}=\Delta C_{1}=0$,
and $\Delta L_{1}$ varies. In other words, these plots show the variation
in the real and imaginary parts of the four eigenvalues as a function
of perturbation in the positive capacitance $C_{1}$ or positive inductance
$L_{1}$ in the PRC.\label{fig:Dispersion}}
\end{figure}

We present an example here to evaluate the analysis presented in the
previous sections. There are many different combinations of values
that will satisfy the EPD condition for the PRC's elements, and here
we assume the following set of values as an example: $L_{1}=1\mathrm{\:\mu H}$,
$L_{0}=-1\mathrm{\:\mu H}$, $C_{1}=0.25\mathrm{\:nF}$ and $C_{0}=-1\mathrm{\:nF}$.
The results in Figure \ref{fig:Dispersion}(a) illustrate the two
branches of the real and imaginary parts of perturbed eigenvalues
calculated from the eigenvalue problem in Equation (\ref{eq:RecipEigValProb}),
varying $C_{1}$ in the neighborhood of $C_{1,\mathrm{e}}=0.25\mathrm{\:nF}$.
In this plot we assume that $\Delta L_{0}=\Delta L_{1}=\Delta C_{0}=0$
and we only perturb $\Delta C_{1}=\left(C_{1}-C_{1,\mathrm{e}}\right)/C_{1,\mathrm{e}}$.
The results in Figure \ref{fig:Dispersion}(b) exhibit the two branches
of the real and imaginary parts of perturbed eigenvalues calculated
from the eigenvalue problem in Equation (\ref{eq:RecipEigValProb}),
varying $L_{1}$ in the proximity of $L_{1,\mathrm{e}}=1\mathrm{\:\mu H}$.
In this plot, we take into account that $\Delta L_{0}=\Delta C_{0}=\Delta C_{1}=0$
and we perturb $\Delta L_{1}=\left(L_{1}-L_{1,\mathrm{e}}\right)/L_{1,\mathrm{e}}$.
We obtain $s_{\mathrm{e}}=\mathrm{i}\omega_{\mathrm{e}}=\mathrm{i}4.47\times10^{7}\:\mathrm{rad/s}$
for this example and the coalesced eigenvalues at EPD are exceedingly
sensitive to perturbations in circuit parameters, i.e., elements values.
In the vicinity of the EPD, we observe a bifurcation, which is one
of the most distinctive features of the EPD \cite{seyranian2005coupling,cartarius2012nonlinear,gutohrlein2013bifurcations,gutohrlein2016bifurcations}.
Although the Taylor series expansion fails in the vicinity of an EPD,
the Newton-Puiseux series can nevertheless be used to conduct a perturbation
analysis \cite[Chapter 2]{seyranian2003multiparameter}.

In order to validate the theoretical results presented in Figures
\ref{fig:Dispersion}(a) and (b), we use a time-domain circuit simulator
powered by Keysight Advanced Design System (Keysight ADS), the most
prestigious software for designing and analyzing circuits. We run
the time-domain simulation to compute the voltages (and therefore
charges) of the capacitors and take a fast Fourier transform (FFT)
of the calculated time-domain results to compute the eigenfrequencies.
The calculated values, i.e., eigenvalues of the circuit $s=\mathrm{i}\omega$,
are illustrated in Figures \ref{fig:Dispersion}(a) and (b) by black
hollow circles. It is worth noting that the results are obtained in
the stable region where the eigenvalues are purely imaginary, i.e.,
the real parts of the eigenvalues are zero. Further, the eigenvalues
are computed only at feasible frequencies, i.e., positive frequencies.
The results of extensive circuit simulations are in excellent agreement
with those obtained from theoretical calculations.

\begin{figure}
\begin{centering}
\includegraphics[width=0.8\textwidth]{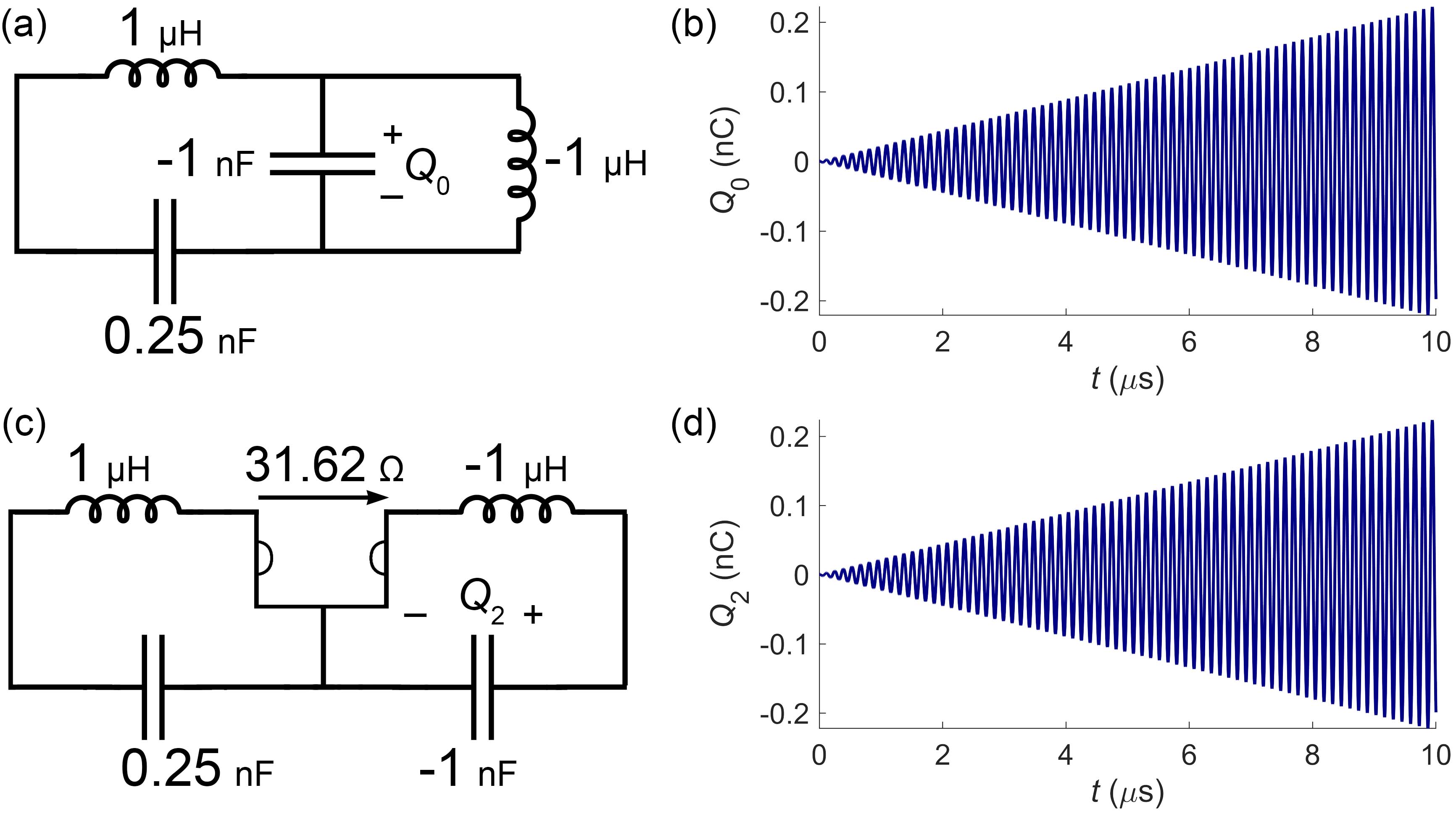}
\par\end{centering}
\caption{Comparison between the time-domain simulation results of (a) the PRC
and (c) the GNC with the proposed equivalent element values. The stored
charges in the capacitor with the negative capacitance in (b) the
PRC $Q_{0}\left(t\right)$ and (d) the GNC $Q_{2}\left(t\right)$
under the EPD condition and by applying the equivalency conditions.\label{fig:CircuitTimeSim}}
\end{figure}

As a next step, we demonstrate that the PRC can be equated with the
GNC under the conditions outlined in Section \ref{sec:Equivalency}.
The PRC with the required values for the elements is shown in \ref{fig:CircuitTimeSim}(a).
The extensive time-domain simulation result generated using the Keysight
ADS circuit simulator is shown in Figure \ref{fig:CircuitTimeSim}(b),
which shows the stored charge $Q_{0}$ in the capacitor with negative
capacitance $C_{0}$. To calculate the charge stored in capacitor
$C_{0}$, we compute the capacitor voltage $V_{C_{0}}$ using Keysight
ADS and calculate the charge through an equation $Q_{0}=C_{0}V_{C_{0}}$.
We should mention that we put $1\:\mathrm{mV}$ as an initial voltage
on $C_{1}$ in the time-domain simulator to establish oscillation.
Figure \ref{fig:CircuitTimeSim}(b) shows that stored charge increase
linearly with time. A significant aspect of the degeneracy of eigenvalues
is that it is the result of coalescing circuit eigenvalues and eigenvectors
that are also associated with a double pole in the circuit. It is
evident from the linear growth over time that a second-order EPD exists
in the circuit.

The GNC with the required values for the elements is shown in \ref{fig:CircuitTimeSim}(c).
It is convenient to consider that counterpart capacitances and inductances
in both circuits have the same value. So, the values of the elements
in the left and right resonators of the GNC are considered equal to
the values used in the previous example for the PRC. Also, the equivalency
value of the gyration resistance $R_{\mathrm{g,eq}}=31.621\mathrm{\:\Omega}$
is calculated based on the equivalency condition discussed in Section
\ref{sec:Equivalency}. The extensive time-domain simulation result
generated using the Keysight ADS circuit simulator is displayed in
Figure \ref{fig:CircuitTimeSim}(d), which represents the stored charge
$Q_{2}$ in the capacitor with the negative capacitance $C_{2}$.
We assigned $1\:\mathrm{mV}$ as the initial voltage on $C_{1}$ in
the time-domain simulator. According to Figure \ref{fig:CircuitTimeSim}(d),
the stored charge grow linearly with increasing time. In addition,
we observe the same behavior on linear growth in Figure \ref{fig:CircuitTimeSim}(b)
and (d), demonstrating the equivalency of eigenfrequencies in the
PRC and GNC.

\section{Conclusions\label{sec:Conclusions}}

We have synthesized a conservative (lossless) electric circuit capable
of attaining a nontrivial Jordan canonical form for its evolution
matrix and consequently exhibiting an EPD. The circuit is composed
of solely conservative reciprocal elements (capacitors and inductors)
and the shared capacitance and parallel inductance should be negative.
Interestingly, we found that the reciprocal and nonreciprocal circuits
presented in our previous papers can produce exactly the same Jordan
canonical form. We also found that the nonreciprocity is manifested
in breakdown of the certain symmetry of the set of eigenvectors as
well as in the Lagrangian. Further, we have thoroughly tested and
confirmed all our significant findings using numerical simulation
using commercial circuit simulator software.

\renewcommand{\sectionname}{Appendix}
\counterwithin{section}{part}
\renewcommand{\thesection}{\Alph{section}}
\setcounter{section}{0}
\renewcommand{\theequation}{\Alph{section}.\arabic{equation}}

\section{Kirchoff's Equations for The PRC\label{sec:Kirchoff's-Equations}}

The following is a concise review of the fundamental equations of
the circuit shown in Figure \ref{fig:RecipCir} based on Kirchhoff's
laws. According to Kirchhoff's voltage law in the two fundamental
loops of the PRC we have

\begin{subequations}
\begin{equation}
\ddot{q}_{1}+\left(\frac{1}{L_{1}C_{1}}+\frac{1}{L_{1}C_{0}}\right)q_{1}-\left(\frac{1}{L_{1}C_{0}}\right)q_{0}=0,
\end{equation}

\begin{equation}
\ddot{q}_{0}-\left(\frac{1}{L_{0}C_{0}}\right)q_{1}+\frac{1}{L_{0}C_{0}}q_{0}=0.
\end{equation}
\end{subequations}
Then, the circuit vector evolution equation and the eigenvalue problem
are expressed as

\begin{equation}
\partial_{t}\mathsf{q}_{\mathrm{r}}=\mathscr{C}\mathsf{q}_{\mathrm{r}},\quad\mathscr{C}_{\mathrm{r}}=\left[\begin{array}{rrcr}
0 & 0 & 1 & 0\\
0 & 0 & 0 & 1\\
-\left(\frac{1}{L_{1}C_{1}}+\frac{1}{L_{1}C_{0}}\right) & \frac{1}{L_{1}C_{0}} & 0 & 0\\
\frac{1}{L_{0}C_{0}} & -\frac{1}{L_{0}C_{0}} & 0 & 0
\end{array}\right],\quad\mathsf{q}_{\mathrm{r}}=\left[\begin{array}{c}
\mathrm{q}_{\mathrm{r}}\\
\partial_{t}\mathrm{q}_{\mathrm{r}}
\end{array}\right],\quad\mathrm{q}_{\mathrm{r}}=\left[\begin{array}{c}
q_{1}\\
q_{0}
\end{array}\right],\label{eq:cirma1a-1}
\end{equation}
which is in full agreement with circuit vector evolution equation
calculated by the Euler-Lagrange formulation in Equation (\ref{eq:RecipEigValProb}).

\section{Basics of Electric Networks\label{sec:Basics-of-Electric}}

As a matter of self-consistency, we provide here an overview of the
basic concepts and notations of electrical network theory. Graph theory
concepts of branches (edges), nodes (vertices) and their incidences
are used in the construction of electrical network theory. The Kirchhoff
current and voltage laws can be used in this approach, which is effective
in loop analysis and selecting independent variables. Specifically,
we are interested in conservative electrical networks, which are composed
of three types of electric elements: capacitors, inductors, and gyrators
\cite{tzeng2006theory}. Inductors and capacitors are two-terminal
electric elements, while gyrators are four-terminal electric elements.

\subsection{Current-voltage relation in the circuit elements\label{subsec:cir-elem}}

\begin{figure}
\centering{}\includegraphics[width=0.35\textwidth]{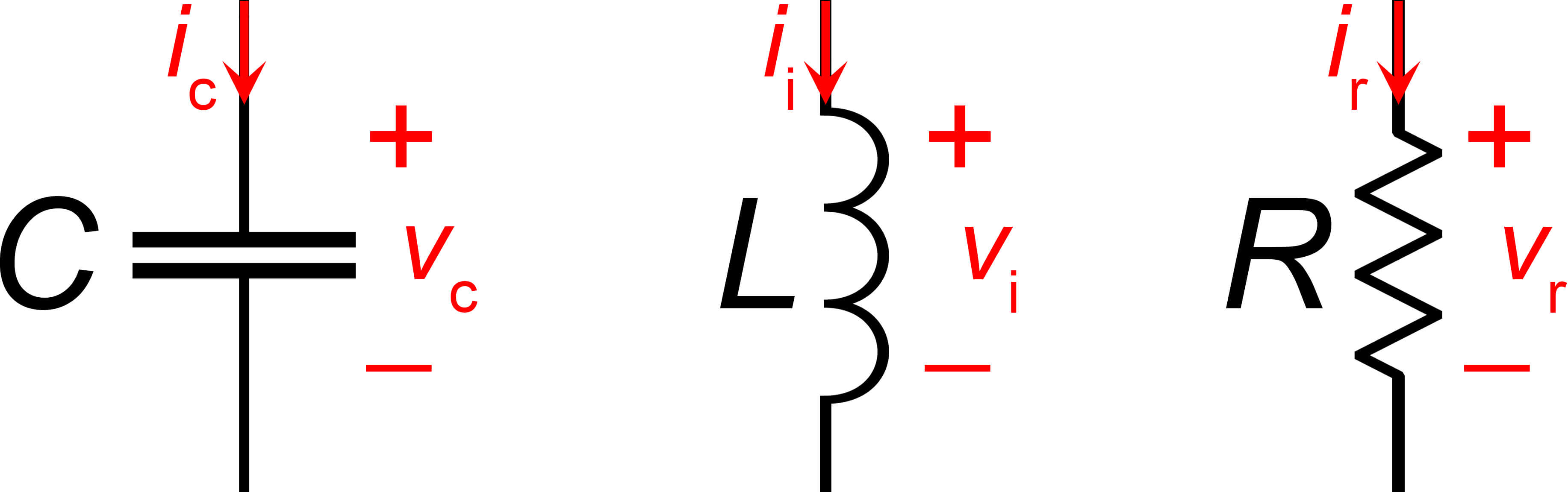}\caption{Capacitance, inductance and resistance symbols and the relevant parameters
in the electric circuits.\label{fig:cir-CLR}}
\end{figure}

In this case, a capacitor, an inductor, a resistor, and a gyrator
are the elements of the basic electric circuit. These elements are
characterized by current-voltage relationships as follows \cite{cauer1958synthesis,balabanian1969electrical,hayt1984electronic,irwin2020basic}:
\begin{equation}
i_{\mathrm{c}}=C\partial_{t}v_{\mathrm{c}},\quad v_{\mathrm{l}}=L\partial_{t}i_{\mathrm{l}},\quad v_{\mathrm{r}}=Ri_{\mathrm{r}},\label{eq:cirvc1a}
\end{equation}
where $i_{\mathrm{c}}$, $i_{\mathrm{l}}$ and $i_{\mathrm{r}}$ are
currents and $v_{\mathrm{c}}$, $v_{\mathrm{l}}$ and $v_{\mathrm{r}}$
are voltages, and real $C$, $L$ and $R$ are called respectively
the capacitance, the inductance and the resistance as shown in Figure
\ref{fig:cir-CLR}.

In addition to the current and voltage we introduce the charge $q$
and the momentum $p$ as

\begin{subequations}
\begin{gather}
q\left(t\right)=\intop i\left(t\right)\,dt,\quad i\left(t\right)=\partial_{t}q,
\end{gather}
\begin{equation}
p\left(t\right)=\intop v\left(t\right)\,dt,\quad v\left(t\right)=\partial_{t}p.
\end{equation}
\end{subequations}
Also, we describe the stored energy parameter $w$ for the elements.
Then, the current-voltage-charge relations, the stored energy and
the Lagrangians associated with the circuit elements are represented
as \cite[Chapter 3]{richards1959manual}

\begin{subequations}
\begin{gather}
\text{Capacitor: }v_{\mathrm{c}}=\frac{q_{\mathrm{c}}}{C},\quad i_{\mathrm{c}}=\partial_{t}q_{\mathrm{c}}=C\partial_{t}v_{\mathrm{c}},\quad q_{\mathrm{c}}=Cv_{\mathrm{c}}=C\partial_{t}p_{\mathrm{c}},
\end{gather}

\begin{equation}
w_{\mathrm{c}}=\frac{1}{2}q_{\mathrm{c}}v_{\mathrm{c}}=\frac{1}{2}\frac{q_{\mathrm{c}}^{2}}{C}=\frac{1}{2}Cv_{\mathrm{c}}^{2}=\frac{1}{2}C\left(\partial_{t}p_{\mathrm{c}}\right)^{2},
\end{equation}
\begin{equation}
\mathcal{L}_{\mathrm{c}}=\frac{q_{\mathrm{c}}^{2}}{2C}.
\end{equation}

\begin{gather}
\text{Inductor: }v_{\mathrm{i}}=L\partial_{t}i_{\mathrm{i}},\quad p_{\mathrm{i}}=Li_{\mathrm{i}}=L\partial_{t}q_{\mathrm{i}},\quad\partial_{t}q_{\mathrm{i}}=\frac{p_{\mathrm{i}}}{L},
\end{gather}
\begin{equation}
w_{\mathrm{i}}=\frac{1}{2}p_{\mathrm{i}}i_{\mathrm{i}}=\frac{1}{2}Li_{\mathrm{i}}^{2}=\frac{1}{2}L\left(\partial_{t}q_{\mathrm{i}}\right)^{2}=\frac{1}{2}\frac{p_{\mathrm{i}}^{2}}{L},
\end{equation}
\begin{equation}
\mathcal{L}_{\mathrm{i}}=\frac{L\left(\partial_{t}q_{\mathrm{i}}\right)^{2}}{2}.
\end{equation}

\begin{equation}
\text{Resistor: }v_{\mathrm{r}}=Ri_{\mathrm{r}},\quad p_{\mathrm{r}}=Rq_{\mathrm{r}}.
\end{equation}
\end{subequations}

\subsection{Negative impedance converter\label{subsec:neg-RCL}}

Our circuits require negative capacitances and inductances, which
can be provided by a number of physical devices. \cite[Chapter 29]{dorf2018engineering}.
Figure \ref{fig:neg-imp} shows suggested circuits that utilize operational
amplifiers (opamps) to obtain negative impedance, capacitance and
inductance \cite[Chapter 10]{izadian2023fundamentals}. The voltage
and current relation and input impedance for circuits depicted in
Figure \ref{fig:neg-imp} are respectively as follows:
\begin{figure}
\begin{centering}
\includegraphics[width=0.77\textwidth]{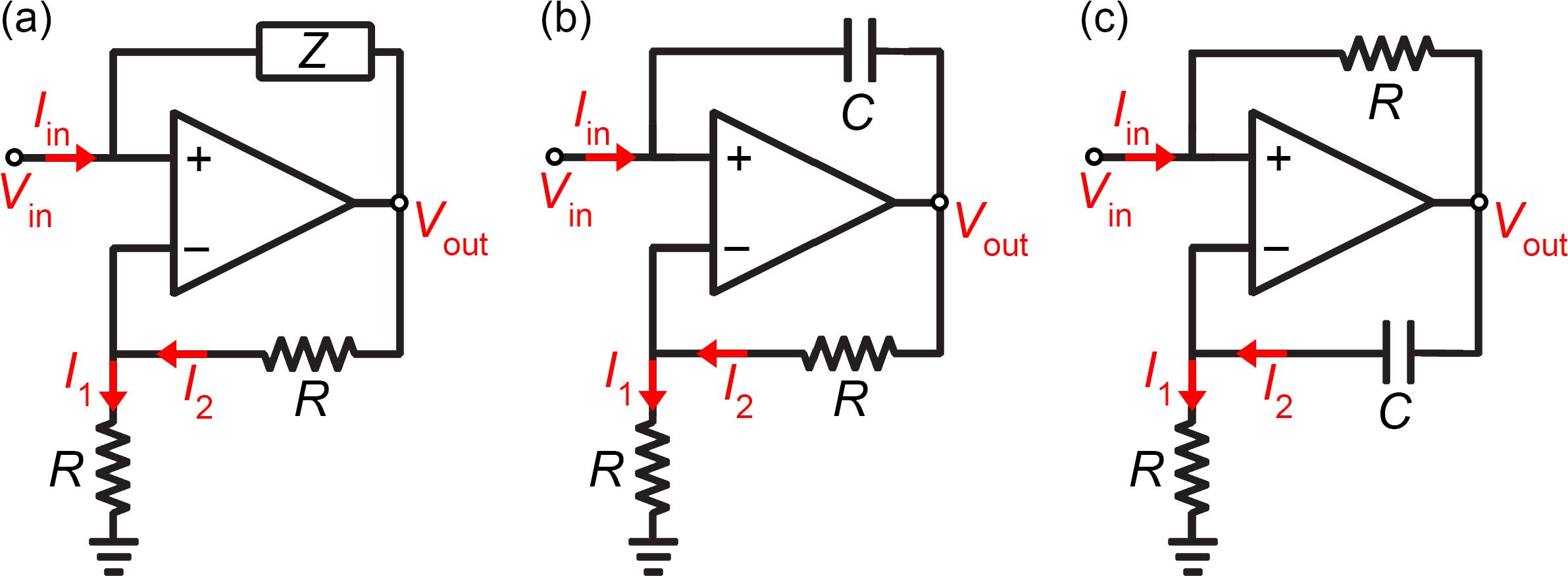}
\par\end{centering}
\centering{}\caption{Opamp-based negative (a) impedance converter, (b) capacitance converter,
(c) inductance converter.\label{fig:neg-imp}}
\end{figure}
(i) for the negative impedance converter (Figure \ref{fig:neg-imp}(a)):

\begin{equation}
I_{1}=I_{2}=\frac{V_{\mathrm{in}}}{R},\quad V_{\mathrm{o}}=2V_{\mathrm{in}}\;\Rightarrow\;Z_{\mathrm{in}}=\frac{V_{\mathrm{in}}}{I_{\mathrm{in}}}=-Z;\label{eq:opamp1a}
\end{equation}
(ii) for the negative capacitance converter(Figure \ref{fig:neg-imp}(b)):

\begin{gather}
I_{1}=I_{2}=\frac{V_{\mathrm{in}}}{R},\quad V_{\mathrm{o}}=2V_{\mathrm{in}}\;\Rightarrow\;Z_{\mathrm{in}}=\frac{V_{\mathrm{in}}}{I_{\mathrm{in}}}=-\frac{1}{\mathrm{i}\omega C};\label{eq:opamp1b}
\end{gather}
(iii) for the negative inductance converter (Figure \ref{fig:neg-imp}(c)):

\begin{gather}
I_{1}=I_{2}=\frac{V_{\mathrm{in}}}{R},\quad V_{\mathrm{o}}=V_{\mathrm{in}}\left(1+\frac{1}{\mathrm{i}\omega RC}\right)\;\Rightarrow\;Z_{\mathrm{in}}=\frac{V_{\mathrm{in}}}{I_{\mathrm{in}}}=-\mathrm{i}\omega\left(R^{2}C\right).\label{eq:opamp1c}
\end{gather}

There are limitations associated with any physical implementation
of an opamp due to deviations from ideal assumptions. Most opamps
deviate from ideal conditions as a result of their limited frequency
band and frequency dependence. A proper tuning of the circuit elements
can restore the EPD property at a single frequency, as the EPD occurs
at a single frequency. It is important to note that the opamp-based
circuits of impedance converters shown in Figure \ref{fig:neg-imp}
are only examples; there are many other implementations with different
features available.

\section{Single LC Circuit Analysis\label{sec:Single-LC-Circuit}}

The circuit matrix of the single LC resonator is given by

\begin{equation}
\mathscr{C}_{\mathrm{LC}}=\left[\begin{array}{rr}
0 & -1\\
\omega_{0}^{2} & 0
\end{array}\right],\quad\omega_{0}=\frac{1}{\sqrt{LC}}
\end{equation}
where $\omega_{0}$ is the resonance frequency of the single LC resonator.
Then, the eigenvalues are expressed as $s_{1,2}=\pm\mathrm{i}\omega_{0}$
and the corresponding eigenvectors by assuming state vector as the
stored charge in the capacitance $q$ and its first derivative, $\Psi_{\mathrm{LC}}=\left[q,\dot{q}\right]$,
are obtained as

\begin{equation}
\mathscr{V}_{\mathrm{LC}}=\left[\begin{array}{rr}
\frac{1}{2} & \frac{1}{2}\\
\mathrm{i}\frac{1}{2\sqrt{LC}} & -\mathrm{i}\frac{1}{2\sqrt{LC}}
\end{array}\right]=\left[\begin{array}{rr}
\frac{1}{2} & \frac{1}{2}\\
\mathrm{i}\frac{\omega_{0}}{2} & -\mathrm{i}\frac{\omega_{0}}{2}
\end{array}\right].
\end{equation}
We define impedance as a ratio between voltage ($v$) and current
($i$). Also, by using the definition of charge in the state vector
which is defined based on the stored charge in the capacitance $q$,
we calculate resonator impedance as a ratio between capacitor voltage
($q/C$) and capacitor current ($\dot{q}$) as

\begin{equation}
Z_{1,2}=\pm\frac{1}{\mathrm{i}\omega_{0}C},\quad\left|Z_{1,2}\right|=\sqrt{\frac{L}{C}},
\end{equation}
which is defined as an impedance of single resonator.

\section{List of Abbreviations\label{sec:AbbList}}
\begin{itemize}
\item EPD: exceptional point of degeneracy
\item GNC: gyrator-based nonreciprocal circuit
\item PRC: principle reciprocal circuit
\end{itemize}
\begin{quotation}
\textbf{\vspace{0.1cm}
}
\end{quotation}
\textbf{Data Availability:} The data that supports the findings of
this study are available within the article.\textbf{\vspace{0.2cm}
}

\textbf{Acknowledgment:} This research was supported by Air Force
Office of Scientific Research (AFOSR) Grant No. FA9550-19-1-0103.

\end{document}